\documentclass[conference,compsoc]{IEEEtran}
\usepackage{amsmath}
\usepackage{amsthm}

\usepackage{cancel}

\usepackage{amssymb}
\usepackage{amsfonts}
\usepackage{utfsym} 
\usepackage[colorlinks=true,urlcolor=black]{hyperref}
\usepackage{hyperref} 
\usepackage{cleveref}
\crefname{section}{Sec.}{Secs.}
\Crefname{section}{Section}{Sections}
\Crefname{table}{Table}{Tables}
\crefname{table}{Tab.}{Tabs.}
\Crefname{figure}{Figure}{Figures}
\crefname{figure}{Fig.}{Figs.}
\Crefname{equation}{Equation}{Equations}
\crefname{equation}{Eq.}{Eqs.}
\crefname{algorithm}{Alg.}{Algs.}
\Crefname{algorithm}{Algorithm}{Algorithms}
\usepackage{url} 
\usepackage{graphicx}
\usepackage{multirow}
\usepackage{makecell}
\usepackage[flushleft]{threeparttable}
\usepackage{colortbl}
\usepackage{booktabs}
\usepackage{xspace}
\usepackage[dvipsnames]{xcolor}
\usepackage{enumitem}
\usepackage[most]{tcolorbox} 
\usepackage{algorithm}
\usepackage{algorithmic}

\tcbset{
  takeaway/.style={
    colback=gray!10,       
    colframe=gray!60,      
    boxrule=0.8pt,         
    arc=4mm,               
    left=5mm, right=5mm,   
    top=3mm, bottom=3mm,   
    fonttitle=\bfseries,
    coltitle=black
  }
}

\ifCLASSOPTIONcompsoc
  \usepackage[nocompress]{cite}
\else
  \usepackage{cite}
\fi
\ifCLASSINFOpdf
\else
\fi
\begin{document}
%
\title{Shared Vulnerabilities in Robustness-Optimized Defenses:\\One Breach Exposes the Family}



%
\author{\IEEEauthorblockN{Hanrui Wang\IEEEauthorrefmark{1},
Ruihao Zheng\IEEEauthorrefmark{2},
Shuo Wang\IEEEauthorrefmark{3}, 
Isao Echizen\IEEEauthorrefmark{1},
Xingbo Dong\IEEEauthorrefmark{2}, and Zhe Jin\IEEEauthorrefmark{2}}
\IEEEauthorblockA{\IEEEauthorrefmark{1}National Institute of Informatics, Tokyo, Japan\\email: \{hanrui\_wang, iechizen\}@nii.ac.jp}
\IEEEauthorblockA{\IEEEauthorrefmark{2}Anhui University, Hefei, China\\email: \{r124302027, xingbo.dong, jinzhe\}@ahu.edu.cn}
\IEEEauthorblockA{\IEEEauthorrefmark{3}Shanghai Jiao Tong University, Shanghai, China\\email: wangshuosj@sjtu.edu.cn}}


\maketitle

\begin{abstract}
Adversarial robustness optimization aims to preserve correct prediction under adversarial perturbations, and has produced substantial robustness gains through methods such as adversarial training and adversarial purification. However, we identify a new security risk: these gains can create shared vulnerabilities across defenses. Once one representative robustness-optimized defense is effectively breached, the broader family may become exposed. Studying this risk requires separating genuine transferability from distortion-induced degradation and from the algorithmic gains of sophisticated attacks. We therefore introduce stricter transfer-only protocols and a deliberately simple adaptive attack, \emph{PGDTransfer}, to test whether robustness-optimized defenses share transfer-only vulnerability under controlled conditions. We further introduce \emph{Adversarial Sensitivity Maps (AdvSMs)} to visualize and quantify shared alignment beyond differentiable classifiers, including stochastic and non-differentiable defenses. Across adversarially trained classifiers, purification-based defenses, and LVLMs with robust visual encoders, we identify natural transferability within each robustness family, i.e., transfer that arises even with simple PGD-style optimization rather than specialized transferable-attack design. The risk is already severe for purification: PGDTransfer reaches an average transfer attack success rate of $80.4\%$ across filtering-, compression-, and diffusion-based purifiers under $\epsilon=4/255$, suggesting that purifier defenses may no longer provide reliable protection. As attacks improve, currently stronger robustness families may face the same risk. Future defenses should therefore treat vulnerability diversity and transfer-only isolation as security objectives, rather than optimizing only individual robustness. \emph{The implementation is available at \url{https://github.com/azrealwang/AdvSM}.}
\end{abstract}


%
\IEEEpeerreviewmaketitle

\section{Introduction}
\label{sec_intro}

Adversarial attacks have raised serious concerns about model robustness, showing that imperceptible perturbations can cause models to misbehave~\cite{goodfellow2014explaining,madry2018towards}. In response, many defenses have been developed to mitigate adversarial threats. Adversarial detection identifies adversarial examples and rejects suspicious inputs~\cite{metzen2017detecting,carlini2017adversarial}. Adversarial training (\Cref{tab_RW_at}) improves robustness by training models on adversarial examples~\cite{shafahi2019adversarial,wongf2020ast}. Adversarial purification (\Cref{tab_RW_purification}) removes or suppresses perturbations before prediction through denoising or stochastic transformations~\cite{nie2022diffusion,lee2023robust}. Although these defenses differ in implementation, adversarial training and purification share a common objective: preserving correct prediction under adversarial perturbations rather than merely rejecting suspicious inputs. We refer to defenses with this objective as \emph{robustness optimization}. Such defenses seek to rely less on non-robust directions and more on task-relevant robust features, leading to substantial reported robustness gains~\cite{liu2025comprehensive,sun2025sample}.

\begin{figure}[!t]
    \centering
    \includegraphics[width=2.8in]{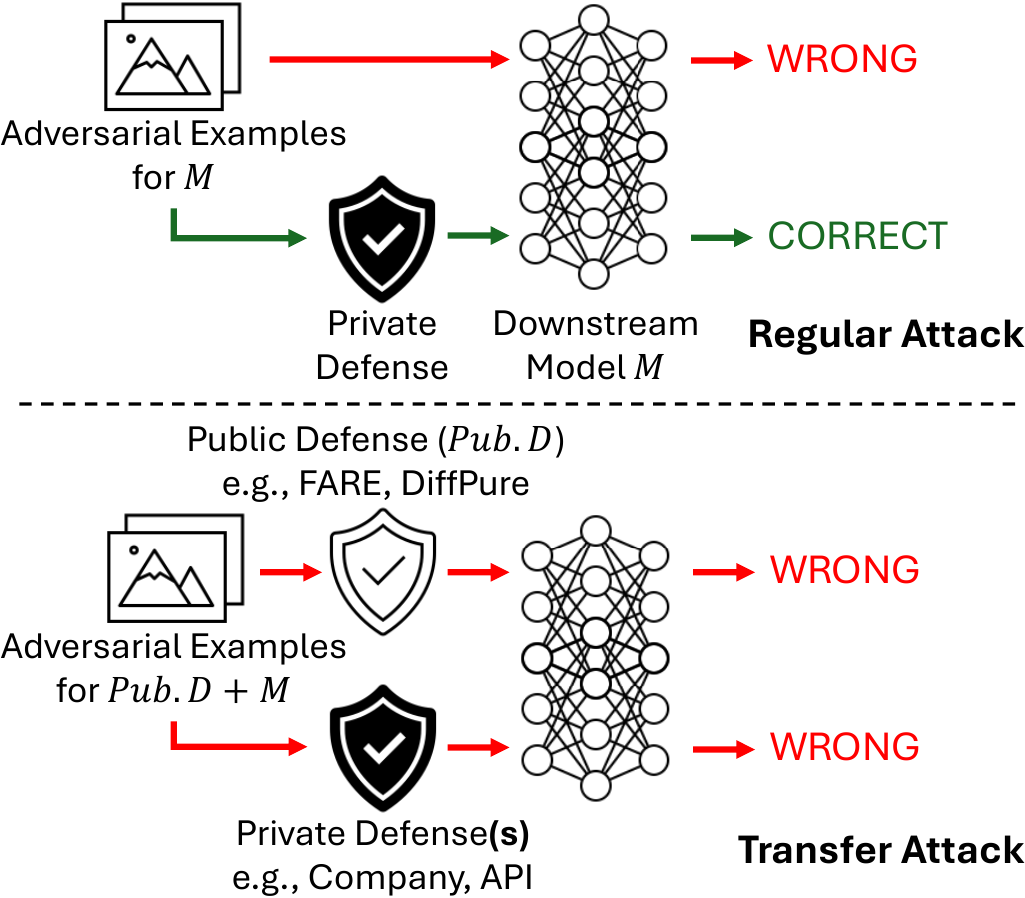}
    \caption{Illustration of defense-isolation risk. A regular attack on the downstream model can be blocked by a private defense. In contrast, a transfer attack crafted through a public representative defense can remain effective against private defenses in the same robustness family, even without target gradients, queries, or adaptation. Thus, one breached representative defense may expose the broader family.}
    \label{fig_risk}
\end{figure}

However, we find that robustness optimization can create shared vulnerabilities across defenses. When optimized toward similar robust behavior, defenses may retain aligned adversarial sensitivities within the same family and task setting. As a result, adversarial examples crafted on one suitable defense can transfer to others without target gradients, queries, or adaptation, as illustrated in \Cref{fig_risk}. This shifts the security concern from isolated robustness to defense isolation: \textbf{once one representative defense is breached, the broader family may become exposed.}

This shared vulnerability raises three deployment-level security concerns.
\textbf{Risk 1: Public representatives may expose private systems.}
Even if deployed defenses are private, their strategies are often public, and representative checkpoints or implementations may exist. Under shared vulnerability, these public representatives may serve as effective surrogates for private systems in the same robustness family. 
\textbf{Risk 2: Defense randomization may fail.} 
Some defenses randomly switch among modules to hide the exact defense used at inference time~\cite{dbouk2023robustness}. If candidate modules share aligned sensitivity, this uncertainty no longer prevents transfer-only attacks: one surrogate adversarial example may remain effective across them. 
\textbf{Risk 3: Currently robust defenses may be compromised by future stronger attacks.}
High individual robustness against current attacks does not remove shared vulnerability. Our results show that a surrogate adaptive attack weakens filtering-, compression-, and diffusion-based purifiers together. The same risk may extend to currently stronger robustness families as future attacks improve.

\begin{table}[!t]
    \centering
    \begin{threeparttable}
        \caption{Adversarial-training-based defenses considered in related work and evaluated in this work.}
        \label{tab_RW_at}
        \setlength{\tabcolsep}{2.4mm}{\begin{tabular}{rcc}
            \toprule
            \textbf{Method}&\textbf{Role}&\textbf{Task}\\
            \midrule
            ResNet-50~\cite{salman2020adversarially} [2020]&Classifier&Classification\\
            ConvNeXt-B~\cite{liu2025comprehensive} [2025]&Classifier&Classification\\
            ViT-B~\cite{mo2022adversarial} [2022]&Classifier&Classification\\
            Swin-B~\cite{mo2022adversarial} [2022]&Classifier&Classification\\
            \midrule
            FARE (ViT-L-14)~\cite{schlarmann2024robust} [2024]&Visual encoder&VQA\\
            TeCoA (ViT-L-14)~\cite{mao2023understanding} [2023]&Visual encoder&VQA\\
            SimCLIP (ViT-L-14)~\cite{hossain2026sim} [2026]&Visual encoder&VQA\\
            \bottomrule
        \end{tabular}}
    \end{threeparttable}
\end{table}

\begin{table}[!t]
    \centering
    \begin{threeparttable}
        \caption{Purification-based adversarial defenses considered in related work and evaluated in this work.}
        \label{tab_RW_purification}
        \setlength{\tabcolsep}{0.6mm}{\begin{tabular}{rccc}
            \toprule
            \textbf{Method}&\textbf{Type}&\textbf{Stochastic}&\textbf{Differentiable}\\
            \midrule
            Mean$^*$~\cite{lee1980digital} [1980]&Filter&$\usym{2717}$&$\usym{2713}$\\
            Gaussian$^*$~\cite{jain1989fundamentals} [1989]&Noise+Filter&$\usym{2713}$&$\usym{2713}$\\
            JPEG$^*$~\cite{wallace1991jpeg} [1991]&Compression&$\usym{2717}$&$\usym{2717}$\\
            DefenseGAN~\cite{samangouei2018defense} [2018]&GAN&$\usym{2713}$&$\usym{2713}$\\
            APE-GAN~\cite{jin2019ape} [2019]&GAN&$\usym{2713}$&$\usym{2713}$\\
            DSM-EBM~\cite{yoon2021adversarial} [2021]&EBM&$\usym{2713}$&Potentially OOM\\
            MCMC-EBM~\cite{hill2021stochastic} [2021]&EBM&$\usym{2713}$&Potentially OOM\\
            SOAP~\cite{shi2021online} [2021]&VAE&$\usym{2713}$&$\usym{2713}$\\
            D-VAE~\cite{yu2024purify} [2024]&VAE&$\usym{2713}$&$\usym{2713}$\\
            DiffPure$^*$~\cite{nie2022diffusion} [2022]&Diffusion&$\usym{2713}$&OOM-prone\\
            DDIM$^*$~\cite{lee2023robust} [2023]&Diffusion&$\usym{2713}$&$\usym{2713}$\\
            MimicDiffusion$^*$~\cite{song2024mimicdiffusion} [2024]&Diffusion&$\usym{2713}$&OOM-prone\\
            ContrastDiff$^*$~\cite{bai2024diffusion} [2024]&Diffusion&$\usym{2713}$&OOM-prone\\
            DifFilter~\cite{chen2024diffilter} [2024]&Diffusion&$\usym{2713}$&OOM-prone\\
            ADBM~\cite{li2025adbm} [2025]&Diffusion&$\usym{2713}$&OOM-prone\\
            DCDefense$^*$~\cite{pei2025divide} [2025]&Diffusion&$\usym{2713}$&OOM-prone\\
            SSNI$^*$~\cite{sun2025sample} [2025]&Diffusion&$\usym{2713}$&OOM-prone\\
            IWMFDiff~\cite{wang2025iterative} [2025]&Diffusion&$\usym{2713}$&$\usym{2717}$\\
            LoRID~\cite{zollicoffer2025lorid} [2025]&Diffusion&$\usym{2713}$&OOM-prone\\
            PuriFlow~\cite{park2025adversarial} [2025]&Diffusion&$\usym{2713}$&OOM-prone\\
            \bottomrule
        \end{tabular}}
        \begin{tablenotes}
            \item $^*$ Denotes the eight target defense methods evaluated in our experiments and one surrogate purifier, DDIM,  used for attack optimization.
        \end{tablenotes} 
    \end{threeparttable}
\end{table}

\begin{table}[!t]
    \centering
    \begin{threeparttable}
        \caption{Transferable adversarial attacks considered in related work and evaluated in this work.}
        \label{tab_RW_transferable_attacks}
        \setlength{\tabcolsep}{10mm}{\begin{tabular}{rc}
            \toprule
            \textbf{Method}&\textbf{Strategy}\\
            \midrule
            MBA$^*$~\cite{li2023making} [2023]&Ensemble-based\\
            FAP~\cite{wang2024boosting} [2024]&Generative-based\\
            BFA~\cite{wang2024improving} [2024]&Advanced objective\\
            P2FA~\cite{liu2025pixel2feature} [2025]&Objective-based\\
            AWT~\cite{chen2025enhancing} [2025]&Model-dependent\\
            OPS~\cite{guo2025boosting} [2025]&Input transformation\\
            MEF~\cite{qiu2026boosting} [2026]&Gradient-based\\
            \bottomrule
        \end{tabular}}
        \begin{tablenotes}
            \item $^*$ Ensemble-based attacks are excluded from evaluation, as they rely on multiple models, whereas our evaluation protocol enforces a single-surrogate setting for fair comparison.
        \end{tablenotes} 
    \end{threeparttable}
\end{table}

\begin{table*}[!t]
    \centering
    \begin{threeparttable}
        \caption{Adaptive adversarial attacks considered in related work and evaluated in this work.}
        \label{tab_RW_adaptive_attacks}
        \setlength{\tabcolsep}{1.4mm}
        \renewcommand{\arraystretch}{1.1}
        \begin{tabular}{>{\raggedleft\arraybackslash}p{3cm} p{6.5cm} p{7.4cm}}
            \toprule
            \textbf{Method}&\textbf{Strategy}&\textbf{Comment}\\
            \midrule
            PGD~\cite{madry2018towards} [2018]&Classifier-only attack&Non-adaptive; easy to defend \\
            \cline{1-3}
            BPDA+EOT~\cite{athalye2018obfuscated} [2018]&Identity-function assumption&Handles non-differentiable purifiers; ineffective for diffusion-based purifiers\\
            \cline{1-3}
            DiffPGD~\cite{xue2023diffusion} [2023]&Gradient-based + DDIM surrogate&No EOT; less effective for stochastic purifiers\\
            \cline{1-3}
            DiffAttack~\cite{kang2023diffattack} [2023]&Gradient-based + EOT + deviated-reconstruction loss + modified DDIM surrogate&Uses intermediate diffusion states; requires modifying diffusion models\\
            \cline{1-3}
            DiffHammer~\cite{wang2024diffhammer} [2024]&Gradient-based + EM loss + DDIM surrogate&Replaces EOT with EM-style loss; still multi-pass purification; less effective than ours\\
            \cline{1-3}
            DiffBreak~\cite{kassis2025diffbreak} [2025]&Gradient-based + EOT + LPIPS loss + DDIM surrogate& Constrains LPIPS but not $\ell_\infty$; less effective under the same LPIPS bound\\
            \midrule
            \textbf{PGDTransfer (Ours)}&Gradient-based + EOT + DDIM surrogate&Simple yet effective for transfer across purifiers\\
            \bottomrule
        \end{tabular}
    \end{threeparttable}
\end{table*}

To study whether robustness optimization induces shared vulnerability and weakens defense isolation, three gaps must be addressed.
\textbf{Gap 1: Existing protocols can overestimate genuine transferability.}
Common transferable-attack evaluations often use untargeted attacks with a relatively large $\ell_\infty$ budget, such as $16/255$.\footnote{\label{foot_transfer}\url{https://github.com/Trustworthy-AI-Group/TransferAttack}} Under this setting, attack success can reflect distortion-induced accuracy degradation rather than shared adversarial directions. 
\textbf{Gap 2: Sophisticated attacks make it difficult to isolate the source of transferability.}
Existing transferable attacks (\Cref{tab_RW_transferable_attacks}) and adaptive attacks (\Cref{tab_RW_adaptive_attacks}) introduce specialized objectives, transformations, surrogate designs, or optimization tricks. These methods can improve attack success rate (ASR), but make it difficult to separate algorithmic gains from natural transferability induced by shared vulnerabilities among robustness-optimized defenses. 
\textbf{Gap 3: Shared vulnerability lacks general evidence beyond attack success.}
High transfer ASR shows that an attack transfers, but does not explain whether transfer is supported by defense similarity. Existing gradient-similarity analysis~\cite{demontis2019adversarial,wang2026minimal} mainly applies to differentiable classifiers and is difficult to extend to stochastic or non-differentiable defenses, such as diffusion purifiers or JPEG compression. A general visualization and measurement tool is therefore needed to test whether high transfer ASR is accompanied by measurable alignment among defenses.

We address these gaps with corresponding contributions:
\begin{itemize}
    \item \textbf{Stricter transfer-only protocols.}
    We introduce evaluation protocols that reduce distortion-driven overestimation, including small-budget untargeted transfer, targeted transfer, robustness-mismatch transfer, and clean-correct evaluation. Under these stricter protocols, transfer success is harder to attribute to generic distortion or pre-existing system errors, making persistent transfer stronger evidence of shared vulnerability.

    \item \textbf{PGDTransfer.}
    We introduce PGDTransfer, a deliberately simple adaptive attack for robustness-optimized defenses. It uses projected gradient descent (PGD)~\cite{madry2018towards}, expectation over transformation (EOT)~\cite{athalye2018synthesizing}, and a suitable surrogate when needed, but avoids specialized transferable-attack mechanisms. Across the evaluated robustness families, including adversarially trained classifiers, adversarial purifiers, and large vision-language models (LVLMs) with robust visual encoders, PGDTransfer achieves non-trivial transfer under the stricter transfer-only setting. This shows that transferability can arise from family-level shared vulnerability rather than sophisticated attack design.

    \item \textbf{Adversarial Sensitivity Maps (AdvSMs).}
    We introduce AdvSMs to visualize and quantify shared alignment among robustness-optimized defenses beyond differentiable classifiers. Using AdvSMs, we show that defenses within the same robustness family exhibit aligned adversarial sensitivity, providing a measurable basis for why one breach can expose the family.
\end{itemize}
Together, these results demonstrate that robustness optimization can improve standalone robustness, while weakening defense isolation across systems optimized toward similar robust behavior.

\section{Transfer-Only Threat Model and Evaluation Protocols}
\label{sec_threat_model}

This section formalizes the transfer-only threat model used throughout this work and defines the evaluation protocols for measuring attack transferability. We clarify several questionable assumptions in existing studies and specify the attacker's knowledge, evaluation metrics, optimization objectives, and perturbation budgets to avoid overestimating genuine transferability.

\subsection{Transfer-Only Threat Model}
\label{sec_transfer_only}

In the transfer-only threat model, transferability measures whether adversarial examples crafted on one system remain effective against another system without target-side adaptation. This setting imposes the following restrictions: \emph{i)} the attacker has white-box access to the surrogate system for full adversarial optimization; \emph{ii)} the attacker has no access to the target system, including target parameters, gradients, or query-based target feedback; and \emph{iii)} each attack uses a single surrogate system rather than an ensemble.

Specifically, let $F_s$ denote a surrogate system and $F_t$ denote a target system. We evaluate three types of robustness-optimized systems. The tested system may be a single model, such as a standard or robust classifier, or a multi-module pipeline, such as a purifier followed by a classifier, or a visual encoder followed by a projector and a language model for visual question answering (VQA). Given a clean input-output pair $(x,y)$, where $y$ denotes the task-specific ground-truth output for the evaluated system, the attacker generates a surrogate adversarial example
\begin{equation}
    x_s^{\mathrm{adv}} = \Pi_{\mathcal{B}_{\epsilon}(x)}
    \left(x + \delta_s\right),
\end{equation}
where $\Pi_{\mathcal{B}_{\epsilon}(x)}$ projects the perturbed input into the allowed $\ell_\infty$ ball around $x$. The perturbation $\delta_s$ is optimized only with respect to the surrogate system $F_s$. The resulting adversarial example is then directly fed into the target system, yielding $F_t(x_s^{\mathrm{adv}})$, without any target-specific optimization. We instantiate this definition in our three experimental systems.

\noindent \textbf{Standard and robust classifiers:} Each system is a single image classifier $f$, i.e.,
\begin{equation}
    F_s(x)=f_s(x),\quad F_t(x)=f_t(x),
    \label{eq_classifier}
\end{equation}
and transferability is evaluated between different classifier architectures or checkpoints.

\noindent \textbf{Purification-based defenses:} Each system is a purifier-classifier pipeline, i.e.,
\begin{equation}
    F_s(x)=f(g_s(x)),\quad F_t(x)=f(g_t(x)),
    \label{eq_purifier}
\end{equation}
where $g$ is the purifier and $f$ is the downstream classifier. To isolate transferability among purifiers, we fix the downstream classifier and vary only the purifier.

\noindent \textbf{VQA systems:} Each system consists of a visual encoder $e$, a projector $p$, and a language model $h$, i.e.,
\begin{equation}
    F_s(x,q)=h(p(e_s(x)),q),\ F_t(x,q)=h(p(e_t(x)),q),
    \label{eq_vqa}
\end{equation}
where $x$ is the image, $q$ is the question. To isolate transferability among visual encoders $e$, we fix the projector and language model and vary only $e$. Transfer is evaluated by replacing $x$ with $x_s^{\mathrm{adv}}$ while keeping $q$ unchanged.

\subsubsection*{Controlled component-level transfer}
For purifier-classifier and VQA pipelines, we intentionally fix the downstream components shared by the surrogate and target systems. This does not relax the transfer-only threat model; rather, it isolates whether the varied robustness-optimized component itself, i.e., the purifier or visual encoder, induces shared adversarial sensitivity. Importantly, shared downstream modules alone do not make transfer inevitable. If they did, transfer would remain high regardless of which purifier or encoder is varied. Instead, our later results show weak transfer when the varied components do not share robust behavior, such as CLIP-to-robust-encoder transfer in \Cref{tab_vqa}, or when weak purifiers are used as surrogates against stronger diffusion purifiers in \Cref{tab_ablation_surrogate}. Transfer becomes stronger only when the varied components expose aligned robustness-induced sensitivity. Thus, our setting is a controlled component-level transfer setting, not target-adaptive or query-based attack evaluation.

\subsection{Evaluation Metrics}
\label{sec_metrics}

We report clean accuracy, white-box ASR (WASR), transfer ASR (TASR), and normalized transfer ASR (nTASR). Clean accuracy is measured on unperturbed inputs using the task-specific correctness criterion of each system. For classification, $y$ is the class label. For VQA, $y$ is the ground-truth answer under the VQA evaluation criterion. For untargeted attacks, the ASR of a target system $F_t$ is defined as
\begin{equation}
    \mathrm{ASR}_{t}^{\mathrm{untargeted}}
    =
    \frac{1}{|\mathcal{D}|}
    \sum_{(x,y)\in\mathcal{D}}
    \mathbb{I}
    \left[
        F_t(x_s^{\mathrm{adv}}) \neq y
    \right].
\end{equation}
For targeted attacks with target output $y^{\mathrm{tar}}$, the ASR is defined as
\begin{equation}
    \mathrm{ASR}_{t}^{\mathrm{targeted}}
    =
    \frac{1}{|\mathcal{D}|}
    \sum_{(x,y)\in\mathcal{D}}
    \mathbb{I}
    \left[
        F_t(x_s^{\mathrm{adv}}) = y^{\mathrm{tar}}
    \right].
\end{equation}
WASR is the ASR measured when the evaluated system itself is used as the surrogate, i.e., adversarial examples are generated and tested on the same system. TASR is the ASR measured when adversarial examples generated on a surrogate system $F_s$ are directly evaluated on a different target system $F_t$, where $F_s \neq F_t$. When comparing transferability across target systems with different white-box robustness, such as robust classifiers, we additionally report nTASR:
\begin{equation}
    \mathrm{nTASR}
    =
    \frac{
        \mathrm{TASR}
    }{
        \mathrm{WASR}_t
    }.
\end{equation}
This metric is useful when an attack has low TASR partly because the target system itself is robust under the same attack setting. We report nTASR only for settings where the corresponding target WASR is non-zero.

\subsection{Two Pitfalls in Commonly Used Protocols}
\label{sec_protocol_problem}

Commonly used transferable-attack protocols typically adopt an untargeted objective, i.e., misclassification, under a relatively large $\ell_\infty$ budget, often $\epsilon=16/255$, and evaluate transfer across classifiers with different architectures or checkpoints~\cite{wang2026devling}\footref{foot_transfer}. However, we find that this protocol can lead to two misleading interpretations.

First, high untargeted TASR under a large perturbation budget, such as $\epsilon=16/255$, may reflect distortion-induced accuracy degradation rather than genuine transferable adversarial directions. A perturbation that disrupts task-relevant image content may reduce accuracy across many models even if the models do not share aligned adversarial sensitivities. In such cases, a high untargeted TASR does not necessarily indicate that the attack has found a shared adversarial direction (see PGD in \Cref{tab_performance_transferable_attacks_ut16}).

Second, transferability is often attributed to architectural similarity, but this explanation is highly conditional. Our results in \Cref{tab_performance_transferable_attacks_architecture} show that models with the same architecture can exhibit weak transferability when one is non-robust and the other is robust. Conversely, robust models can exhibit stronger transferability even across different architectures (see \Cref{fig_matrix_classifiers}).

\subsection{Stricter Protocols for Genuine Transferability}
\label{sec_stricter_protocols}

The above pitfalls motivate a stricter evaluation design for reducing distortion-driven overestimation, objective-driven overestimation, and architecture-based misattribution. We therefore adopt four complementary controls.

\noindent \textbf{Small-budget untargeted transfer.}
We regard this as the primary protocol. It uses untargeted attacks under a smaller $\ell_\infty$ budget, $\epsilon=4/255$. This setting reduces the chance that attack success is mainly caused by visible or semantically disruptive distortion.

\noindent \textbf{Targeted transfer.}
We further evaluate targeted transfer under $\epsilon=16/255$. Targeted attacks require the adversarial example to induce a specific target label rather than merely cause any incorrect prediction. This objective makes transferability harder to achieve through generic distortion alone. We additionally report targeted transfer under $\epsilon=4/255$ as an overly strict stress test, but do not recommend it as a primary protocol because most attacks achieve nearly 0\% TASR under this setting (see \Cref{tab_performance_transferable_attacks_nonUT16}).

\noindent \textbf{Robustness-mismatch transfer.}
We also evaluate transfer between non-robust and robust models that share the same architecture. This protocol helps distinguish architecture-driven transfer from robustness-optimization-driven transfer.

\noindent \textbf{Clean-correct subset.}
We evaluate all attacks on a common clean-correct subset, where an input is included only if it is correctly classified or answered by all systems under comparison before attack. This protocol avoids counting pre-existing clean errors as attack success.

\subsection{Scope}
\label{sec_scope}

First, we do not aim to design the strongest transferable attack for non-robust models. Our claim is that transferability can naturally arise when models are optimized toward similar robust behavior; even simple PGD-style attacks may then transfer across defenses, indicating a family-level risk rather than a benefit of sophisticated attack design.

Second, we study transferability induced by robustness optimization, not adversarial detection. Our scope includes robust classifiers, purification-based defenses, and LVLMs with robust visual encoders. Detection behaves like an additional standard binary classifier, and is not expected to induce transferability in the same way. 

Third, we do not study transfer across defense families, such as from adversarial training to purification. We focus on systems within the same defense family and task setting, where they are optimized toward similar robust behavior.

\section{PGDTransfer Attack}
\label{sec_pgdtransfer}

We now instantiate PGDTransfer, a simple adaptive PGD-style attack used to evaluate whether AdvSM-aligned robustness-optimized systems also share adversarial examples. PGDTransfer is intentionally minimal, in order to provide a controlled test of whether transferability can arise under a basic, non-transfer-oriented adaptive optimization procedure. The attack is applied to the systems introduced in \Cref{eq_classifier,eq_purifier,eq_vqa}. When applied to classifiers alone, PGDTransfer reduces to the standard PGD attack~\cite{madry2018towards}.

\subsection{Objective}
\label{sec_pgdtransfer_objective}

PGDTransfer optimizes adversarial examples under the transfer-only threat model defined in \Cref{sec_transfer_only}. Given a clean image $x$ and attack target $y^*$, PGDTransfer optimizes an EOT-averaged task loss:
\begin{equation}
    \mathcal{J}(x')
    =
    \frac{1}{K}
    \sum_{k=1}^{K}
    \mathcal{L}_{\mathrm{task}}
    \left(
    F_s, x', y^*; \omega_k
    \right),
    \label{eq_pgdtransfer_objective}
\end{equation}
where $\mathcal{L}_{\mathrm{task}}(F_s,x',y^*;\omega_k)$ denotes the system-level loss obtained on the surrogate $F_s$. For classification, the surrogate forward pass is $F_s(x';\omega_k)$. For VQA, it is $F_s(x',q;\omega_k)$, where the original question $q$ is fixed and only the image is perturbed. $K$ is the number of EOT samples, and $\omega_k$ captures surrogate stochasticity; for deterministic systems, $K=1$ and $\omega_k$ can be omitted.

The task loss is instantiated by task type. For classification, including purifier-classifier pipelines, $\mathcal{L}_{\mathrm{task}}$ is the standard cross-entropy loss over class labels:
\begin{equation}
    \mathcal{L}_{\mathrm{task}}
    \left(
    F_s,x',y^*
    \right)
    =
    \mathrm{CE}
    \left(
    F_s(x'),y^*
    \right).
\end{equation}
For VQA, $y^*$ denotes an answer sequence with tokens
$y^*=(a_1^*,\ldots,a_R^*)$, and $\mathcal{L}_{\mathrm{task}}$ is the teacher-forced answer negative log-likelihood under the surrogate distribution $P_s$ induced by $F_s(x',q)$:
\begin{equation}
    \mathcal{L}_{\mathrm{task}}
    \left(
    F_s,x',y^*
    \right)
    =
    -
    \sum_{r=1}^{R}
    \log
    P_s
    \left(
    a_r^*
    \mid
    a_{<r}^*, x', q
    \right).
    \label{eq_vqa_loss}
\end{equation}

For untargeted attacks, $y^*=y$ and PGDTransfer maximizes $\mathcal{J}$, where $y$ is the class label for classification or the ground-truth answer sequence for VQA. For targeted attacks, $y^*=y^{\mathrm{tar}}$ and PGDTransfer maximizes $-\mathcal{J}$.

\begin{algorithm}[t]
\caption{PGDTransfer}
\label{alg_pgdtransfer}
\begin{algorithmic}[1]
\REQUIRE Clean image $x$, optional VQA question $q$, attack target $y^*$, surrogate system $F_s$, budget $\epsilon$, step size $\alpha$, iterations $T$, EOT samples $K$, direction $\rho\in\{+1,-1\}$
\ENSURE Surrogate adversarial image $x_s^{\mathrm{adv}}$
\STATE Initialize $x'_0$ randomly in $\mathcal{B}_{\epsilon}(x)$
\STATE $x'_0 \leftarrow \mathrm{clip}(x'_0,0,1)$
\STATE $x_{\mathrm{best}} \leftarrow x'_0$
\STATE $\widetilde{\mathcal{J}}_{\mathrm{best}} \leftarrow -\infty$
\FOR{$t=0$ to $T-1$}
    \STATE $\mathcal{J} \leftarrow 0$
    \FOR{$k=1$ to $K$}
        \STATE Sample stochasticity $\omega_k$
        \STATE $\mathcal{J} \leftarrow \mathcal{J} + \mathcal{L}_{\mathrm{task}}(F_s,x'_t,y^*;\omega_k)/K$
    \ENDFOR
    \STATE $\widetilde{\mathcal{J}} \leftarrow \rho \mathcal{J}$ \COMMENT{$\rho=+1$ for untargeted, $\rho=-1$ for targeted}
    \STATE $g_t \leftarrow \nabla_{x'_t}\widetilde{\mathcal{J}}$
    \IF{$\widetilde{\mathcal{J}} > \widetilde{\mathcal{J}}_{\mathrm{best}}$}
        \STATE $x_{\mathrm{best}} \leftarrow x'_t$
        \STATE $\widetilde{\mathcal{J}}_{\mathrm{best}} \leftarrow \widetilde{\mathcal{J}}$
    \ENDIF
    \STATE $x'_{t+1} \leftarrow x'_t + \alpha\cdot \mathrm{sign}(g_t)$
    \STATE $x'_{t+1} \leftarrow \Pi_{\mathcal{B}_{\epsilon}(x)}(x'_{t+1})$
    \STATE $x'_{t+1} \leftarrow \mathrm{clip}(x'_{t+1},0,1)$
\ENDFOR
\STATE $x_s^{\mathrm{adv}} \leftarrow x_{\mathrm{best}}$
\RETURN $x_s^{\mathrm{adv}}$
\end{algorithmic}
\end{algorithm}

\subsection{Optimization}
\label{sec_pgdtransfer_optimization}

Starting from a random point $x'_0$ within the perturbation ball around $x$, PGDTransfer performs standard projected gradient updates on the surrogate objective. At iteration $t$, we first convert the task loss into the attack objective
\begin{equation}
    \widetilde{\mathcal{J}}(x'_t)
    =
    \rho \mathcal{J}(x'_t),
    \quad
    \rho =
    \begin{cases}
        +1, & \text{untargeted attack},\\
        -1, & \text{targeted attack}.
    \end{cases}
\end{equation}
PGDTransfer then estimates the gradient of this objective with respect to the image input:
\begin{equation}
    g_t
    =
    \nabla_{x'_t}
    \widetilde{\mathcal{J}}(x'_t).
\end{equation}
For purification-based defenses, this gradient is computed through the purifier and the downstream classifier. For VQA systems, it is computed through the visual encoder, multimodal projector, and answer loss, following the surrogate definitions in \Cref{eq_purifier,eq_vqa}.

For the $\ell_\infty$ setting, the update is
\begin{equation}
    x'_{t+1}
    =
    \Pi_{\mathcal{B}_{\epsilon}(x)}
    \left(
    x'_t + \alpha \cdot \mathrm{sign}(g_t)
    \right),
\end{equation}
followed by clipping to the valid range $[0,1]$. We keep the iterate with the highest attack objective and return it as $x_s^{\mathrm{adv}}$.

The optimization is performed only on the surrogate system $F_s$; target systems are used only for transfer evaluation, never for gradient computation, queries, or adaptation. PGDTransfer therefore remains a standard PGD-style attack on a chosen surrogate, with EOT used only to average stochastic surrogate forward passes. This keeps the procedure consistent with the transfer-only threat model. The full procedure is summarized in \Cref{alg_pgdtransfer}.

\begin{figure*}[!t]
    \centering
    \includegraphics[width=\linewidth]{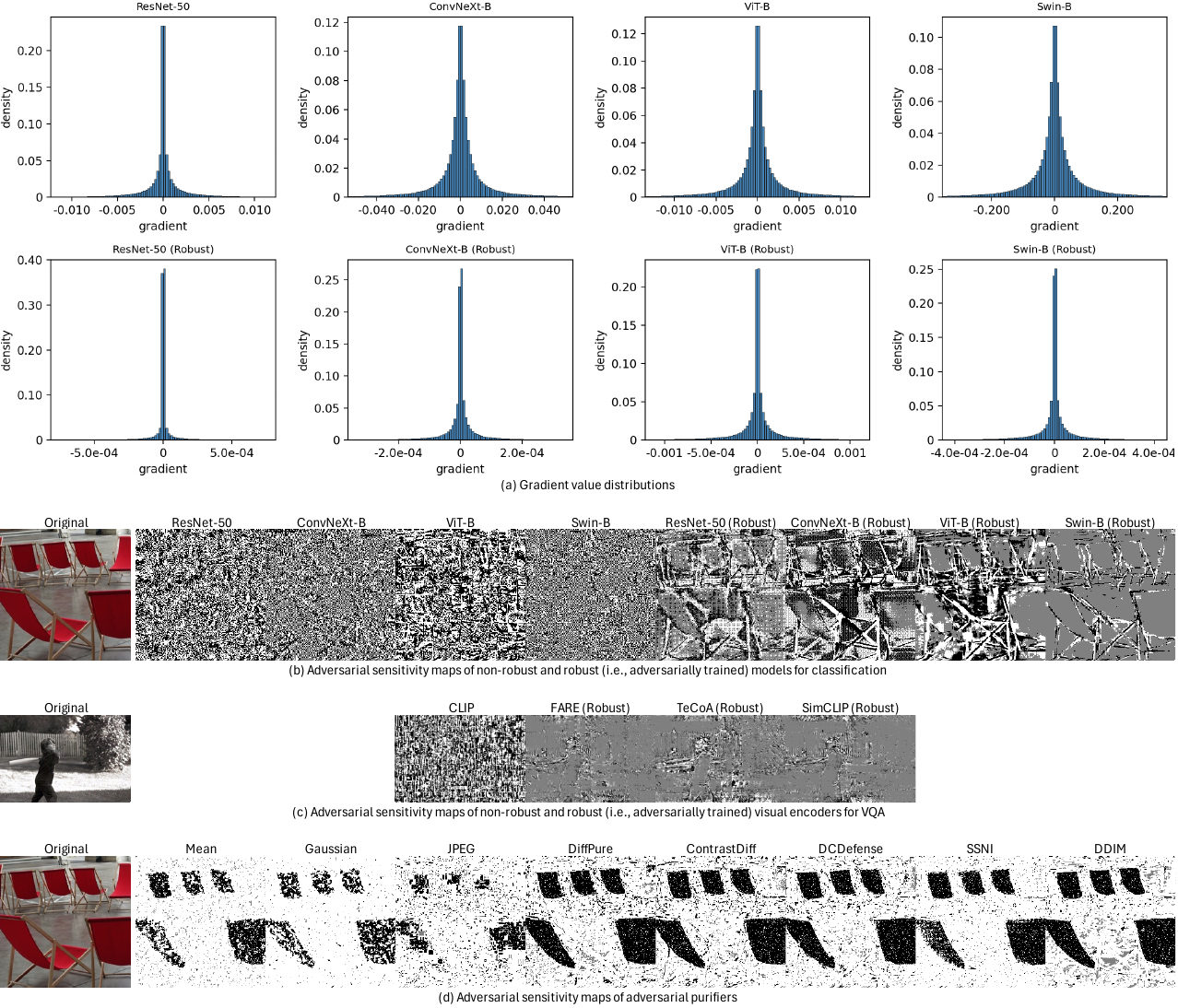}
    \caption{Adversarial sensitivity maps (AdvSMs) reveal shared sensitivity induced by robustness optimization. Compared with non-robust classifiers, robust classifiers exhibit gradients that are more concentrated around zero. Moreover, robustness-optimized defenses show more semantically aligned sensitivity regions. These patterns suggest that shared adversarial sensitivity can emerge within the same robustness family.}
    \label{fig_asm}
\end{figure*}

\section{Adversarial Sensitivity Maps (AdvSMs)}
\label{sec_advsm}

This section introduces AdvSMs, a tool for visualizing and quantifying shared adversarial sensitivity in robustness-optimized defenses. We first review gradient similarity for differentiable classifiers, then define AdvSMs for adversarial training and purification.

\subsection{Preliminary: Model Gradient Similarity}
\label{sec_gradient_similarity}

Prior work has connected attack transferability to gradient alignment: adversarial examples are more likely to transfer when different models have positively aligned loss gradients~\cite{demontis2019adversarial,wang2026minimal}. Following these works, for a classifier $f_k$ and a clean image-label pair
$(x_t,y_t)$, we compute
\begin{equation}
    g_t^{(k)}
    =
    \nabla_{x_t}
    \mathcal{L}\left(f_k(x_t),y_t\right),
\end{equation}
where $\mathcal{L}$ is the cross-entropy loss. We flatten the gradient into a vector and normalize it to remove scale effects:
\begin{equation}
    \hat{g}_t^{(k)}
    =
    \frac{\mathrm{vec}(g_t^{(k)})}
    {\|\mathrm{vec}(g_t^{(k)})\|_2}.
\end{equation}
The gradient similarity between two classifiers $f_i$ and $f_j$ is computed as the
mean cosine similarity over an evaluation subset $\mathcal{D}_{\mathrm{sub}}$:
\begin{equation}
    S_{ij}^{\mathrm{grad}}
    =
    \frac{1}{|\mathcal{D}_{\mathrm{sub}}|}
    \sum_{(x_t,y_t)\in\mathcal{D}_{\mathrm{sub}}}
    \left\langle
    \hat{g}_t^{(i)},\hat{g}_t^{(j)}
    \right\rangle .
\end{equation}

This formulation is useful for differentiable classifiers, but it has two limitations in our setting. First, it cannot be applied to non-differentiable defenses such as JPEG compression, and becomes unstable for stochastic pipelines. Second, cosine similarity over full gradients treats all pixels as active contributors, even though robustness-optimized defenses suppress many less-sensitive regions. As shown in \Cref{fig_asm}(a), robust classifiers produce gradient distributions that are much more concentrated around zero than non-robust classifiers, suggesting that many pixels are effectively ignored by robust feature extraction. Therefore, we need a representation that emphasizes whether a pixel is adversarially usable, rather than its exact gradient magnitude.

\subsection{AdvSMs for Adversarial Training}
\label{sec_advsm_at}

For differentiable systems, AdvSMs convert raw input gradients into ternary pixel-wise sensitivity maps. Here, $F_k$ denotes the $k$-th differentiable system: for image classification, $F_k(x)=f_k(x)$; for VQA, $F_k(x,q)=h(p(e_k(x)),q)$, where only the visual encoder $e_k$ varies. Unlike gradient similarity, AdvSMs do not compare precise gradient values. Instead, they retain only whether each pixel-channel location is positively sensitive, negatively sensitive, or insensitive.

Given the raw input gradient $g_t^{(k)}$ of system $F_k$ on image sample $x_t$, we reshape it into $G_t^{(k)}\in\mathbb{R}^{C\times H\times W}$. For a threshold $\tau$, the AdvSM label at channel $c$ and pixel $(h,w)$ is defined as
\begin{equation}
    \ell_{c,h,w}^{(k,t)}
    =
    \begin{cases}
        0,  & \left|G_{c,h,w}^{(k,t)}\right| \leq \tau,\\
        +1, & \left|G_{c,h,w}^{(k,t)}\right| > \tau
              \ \mathrm{and}\ G_{c,h,w}^{(k,t)} > 0,\\
        -1, & \left|G_{c,h,w}^{(k,t)}\right| > \tau
              \ \mathrm{and}\ G_{c,h,w}^{(k,t)} < 0.
    \end{cases}
\end{equation}
This gives a ternary map
\begin{equation}
    L_t^{(k)} \in \{-1,0,+1\}^{C\times H\times W},
\end{equation}
where $0$ denotes an insensitive location, $+1$ denotes a positively sensitive location, and $-1$ denotes a negatively sensitive location.

To quantify cross-system AdvSM alignment, we flatten two maps and compute their cosine similarity:
\begin{equation}
    \mathrm{sim}
    \left(
    L_t^{(i)},L_t^{(j)}
    \right)
    =
    \frac{
    \mathrm{vec}(L_t^{(i)})^\top
    \mathrm{vec}(L_t^{(j)})
    }{
    \|\mathrm{vec}(L_t^{(i)})\|_2
    \|\mathrm{vec}(L_t^{(j)})\|_2
    }.
\end{equation}
If either map has zero norm, we set the similarity to $0$. The AdvSM similarity between two systems is computed as the mean cosine similarity over an evaluation subset $\mathcal{D}_{\mathrm{sub}}$:
\begin{equation}
    S_{ij}^{\mathrm{at}}
    =
    \frac{1}{|\mathcal{D}_{\mathrm{sub}}|}
    \sum_{x_t\in\mathcal{D}_{\mathrm{sub}}}
    \mathrm{sim}
    \left(
    L_t^{(i)},L_t^{(j)}
    \right),
    \ 
    S_{ii}^{\mathrm{at}}=1.
\end{equation}

As shown in \Cref{fig_asm}(b) and \Cref{fig_asm}(c), non-robust classifiers and visual encoders produce noisy AdvSMs with weak spatial agreement. In contrast, robust classifiers and visual encoders produce maps that concentrate on semantic image regions. This pattern indicates that robustness optimization can align the locations that remain adversarially sensitive, even across different architectures.

\subsection{AdvSMs for Adversarial Purification}
\label{sec_advsm_purification}

Gradient-based maps are not sufficient for adversarial purifiers because many purifiers are stochastic, non-differentiable, or expensive to backpropagate through. We therefore define purifier AdvSMs through output response rather than input gradients. The goal is to measure how each pixel-channel location of the purified output reacts to small random input perturbations.

Let $g_d$ be the $d$-th purifier. For an input $x_t$, we first compute its baseline purified output. For stochastic purifiers, we average over $N$ independent purification runs:
\begin{equation}
    y_0^{(d)}
    =
    \frac{1}{N}
    \sum_{n=1}^{N}
    g_d(x_t;\omega_n),
\end{equation}
where $\omega_n$ denotes the stochasticity of the $n$-th purification run and can be omitted for deterministic purifiers. We then draw $M$ random sign noise tensors $s_m\in\{-1,+1\}^{C\times H\times W}$ and construct
\begin{equation}
    x_t^{(m)}
    =
    \mathrm{clip}
    \left(
    x_t + \epsilon s_m, 0, 1
    \right),
    \quad m=1,\ldots,M.
\end{equation}
For each perturbed input, we again average the purifier output over $N$ independent runs:
\begin{equation}
    \bar{y}_m^{(d)}
    =
    \frac{1}{N}
    \sum_{n=1}^{N}
    g_d(x_t^{(m)};\omega_{m,n}).
\end{equation}
The purifier response is
\begin{equation}
    \delta_m^{(d)}
    =
    \bar{y}_m^{(d)} - y_0^{(d)}.
\end{equation}

For each channel and pixel location, we assign one of three AdvSM labels using a response threshold $\theta$:
\begin{equation}
    \ell_{c,h,w}^{(d,t)}
    =
    \begin{cases}
        0,  & |\delta_{m,c,h,w}^{(d)}| \leq \theta,\ \forall m,\\
        -1, & |\delta_{m,c,h,w}^{(d)}| > \theta,\ \forall m
              \ \mathrm{and}\ 
              \mathrm{ConsSign}_{c,h,w}^{(d,t)},\\
        +1, & \mathrm{otherwise}.
    \end{cases}
\end{equation}
Here, $\mathrm{ConsSign}_{c,h,w}^{(d,t)}$ means that all response $\{\delta_{m,c,h,w}^{(d)}\}_{m=1}^{M}$ have the same sign, while the sampled input noise signs $\{s_{m,c,h,w}\}_{m=1}^{M}$ contain both $+1$ and $-1$.
Thus, $0$ denotes a purified location, and $-1$ denotes a smooth location with a stable one-directional purifier response; in both regions, local adversarial perturbations are suppressed or redirected. In contrast, $+1$ denotes a sensitive location whose purified output remains perturbation-dependent or unstable, and therefore remains usable for adversarial optimization. This produces a purifier AdvSM
\begin{equation}
    L_t^{(d)} \in \{-1,0,+1\}^{C\times H\times W}.
\end{equation}

Cross-purifier similarity is computed with the same ternary-map cosine used for adversarially trained models:
\begin{equation}
    S_{ij}^{\mathrm{pur}}
    =
    \frac{1}{|\mathcal{D}_{\mathrm{sub}}|}
    \sum_{x_t\in\mathcal{D}_{\mathrm{sub}}}
    \mathrm{sim}
    \left(
    L_t^{(i)},L_t^{(j)}
    \right),
    \quad
    S_{ii}^{\mathrm{pur}}=1.
\end{equation}

As shown in \Cref{fig_asm}(d), different purification-based defenses, including filtering, compression, and diffusion-based methods, produce aligned sensitive regions. Their main difference lies in how many pixels are suppressed or smoothed. Stronger purifiers tend to mark more regions as smooth or purified, leaving fewer locations available for adversarial optimization.

\subsection{Choosing a Proper Surrogate}
\label{sec_proper_surrogate}

AdvSMs also provide a practical criterion for selecting a surrogate system. A useful surrogate should expose a compact set of sensitive pixels, so that perturbations optimized on the surrogate are more likely to fall within sensitive regions preserved by other systems in the same robustness family. If a surrogate has too many sensitive pixels, the optimized perturbation may rely on locations that are later purified, smoothed, or ignored by the target. In this case, transferring the adversarial example to the target is similar to passing it through an additional defense layer, where much of the perturbation can be canceled (see \Cref{sec_surrogate_selection}).

\section{Evaluation}
\label{sec_performance}

\subsection{Experimental Settings}
\label{sec_settings}

We evaluate attack transferability in three settings: simple classifier pipelines, purifier-classifier pipelines, and LVLM-based VQA pipelines. Unless otherwise specified, each evaluation uses 500 clean-correct samples, i.e., samples correctly classified or answered by all surrogate and target systems before attack. AdvSM similarity is computed on 100 clean-correct samples.

\noindent \textbf{Datasets.}
We use the ImageNet-compatible dataset from the NIPS 2017 adversarial defense challenge~\cite{nips2017} for classification and purification experiments, and VQAv2~\cite{lin2014microsoft,antol2015vqa} for VQA experiments.

\noindent \textbf{Systems.}
For classifier transfer, each system follows $F(x)=f(x)$. We evaluate four non-robust classifiers, ResNet-50, ConvNeXt-B, ViT-B, and Swin-B, and four RobustBench robust classifiers~\cite{croce2021robustbench}\footnote{\url{https://github.com/RobustBench/robustbench}}, as listed in \Cref{tab_RW_at}. 
For purification transfer, each system follows $F(x)=f(g(x))$: the downstream classifier $f$ is fixed to a non-robust ResNet-50, while the purifier $g$ varies. Target and surrogate purifier settings are listed in \Cref{tab_settings_purification}. 
For VQA transfer, each system follows $F(x,q)=h(p(e(x)),q)$, where $h$ is LLaVA-1.5-7B~\cite{liu2023visual}, $p$ is the multimodal projector, and only the visual encoder $e$ varies. We use CLIP ViT-L/14 as the non-robust visual encoder and the robust ViT-L/14 encoders in \Cref{tab_RW_at}.

\noindent \textbf{Baselines.}
For classifier transfer, we evaluate the single-surrogate attacks in \Cref{tab_RW_transferable_attacks} using TransferAttack~\cite{wang2026devling}\footref{foot_transfer}. For purification transfer, we compare PGDTransfer with the adaptive attacks in \Cref{tab_RW_adaptive_attacks}.

\noindent \textbf{AdvSM configurations.}
For classifier and VQA, we use a gradient threshold of $10^{-5}$ to obtain ternary sensitivity maps. For purifier AdvSMs, we use random sign perturbations with $\epsilon=16/255$, response threshold $\theta=2/255$, $M=5$ perturbation samples, and $N=10$ purification runs per baseline or perturbed input.

\begin{table}[!t]
    \centering
    \begin{threeparttable}
        \caption{Settings of adversarial purifiers.}
        \label{tab_settings_purification}
        \setlength{\tabcolsep}{4.5mm}
        \begin{tabular}{ll}
            \toprule
            Purifier&Settings\\
            \midrule
            Mean&kernel=5\\
            Gaussian&noise std=0.015, kernel=5, sigma=1.5\\
            JPEG&quality=20\%\\
            DiffPure&time\_steps=150\\
            MimicDiffusion&time\_steps=1000, denoise\_steps=100\\
            ContrastDiff&time\_steps=150, sample\_steps=1\\
            SSNI&time\_steps=150, denoise\_steps=150\\
            DCDefense&time\_steps=150, forward\_noise\_steps=1,\\&strength\_l=0.2, strength\_s=0.1\\
            \midrule
            DDIM (surrogate)&time\_steps=150, denoise\_steps=3\\
            \bottomrule
        \end{tabular}
    \end{threeparttable}
\end{table}

\begin{table}[!t]
    \centering
    \begin{threeparttable}
        \caption{Attack configurations.}
        \label{tab_attack_settings}
        \setlength{\tabcolsep}{1.1mm}
        \renewcommand{\arraystretch}{1.08}
        \begin{tabular}{lccccc}
            \toprule
            \textbf{Setting} & \textbf{Objective} & \textbf{$\epsilon$} & \textbf{$\alpha$} & \textbf{$T$} & \textbf{EOT}\\
            \midrule
            Classifier 
            & Untargeted/Targeted 
            & $4/255$ or $16/255$ 
            & $\epsilon/4$ 
            & 10 
            & -- \\
            Purifier 
            & Untargeted/Targeted 
            & $4/255$ or $16/255$ 
            & $1/255$ 
            & 40 
            & 5 \\
            VQA 
            & Untargeted 
            & $4/255$ or $16/255$ 
            & $1/255$ 
            & 100 
            & -- \\
            \bottomrule
        \end{tabular}
    \end{threeparttable}
\end{table}

\begin{table}[!t]
    \centering
    \begin{threeparttable}
        \caption{ASR (\%) of transferable attacks under the commonly used protocol: untargeted transfer with $\ell_\infty$ budget $\epsilon=16/255$. Higher is better.}
        \label{tab_performance_transferable_attacks_ut16}
        \setlength{\tabcolsep}{0.7mm}
        \begin{tabular}{lc|cccc|c}
            \toprule
            \multirow{2}{*}{Attack}&Surrogate&\multicolumn{5}{c}{Target Model}\\
            &Model&ResNet-50&ConvNeXt-B&ViT-B&Swin-B&Avg.\\
            \midrule
            PGD&\multirow{7}{*}{ResNet-50}&\cellcolor{gray!20}100.0&8.8&3.0&6.8&6.2\\
            FAP&&\cellcolor{gray!20}99.8&24.6&11.2&15.4&17.1\\
            BFA&&\cellcolor{gray!20}100.0&71.4&30.6&52.2&51.4\\
            P2FA&&\cellcolor{gray!20}100.0&41.0&16.2&33.4&30.2\\
            AWT&&\cellcolor{gray!20}100.0&65.0&39.2&46.4&50.2\\
            OPS&&\cellcolor{gray!20}99.8&83.4&65.0&67.2&71.9\\
            MEF&&\cellcolor{gray!20}100.0&70.6&33.8&43.8&49.4\\
            \midrule
            PGD&Swin-B (Robust)&79.0&68.0&71.4&95.6&78.5\\
            OPS&ConvNeXt-B&97.2&\cellcolor{gray!20}100.0&92.8&96.6&95.5\\
            \bottomrule
        \end{tabular}
        \begin{tablenotes}
            \item Gray cells indicate WASR. Avg. denotes the average TASR.
            \item The first seven rows follow the common single-surrogate setting with a non-robust ResNet-50 surrogate. The last two rows show that high TASR can also arise from surrogate choice under the same permissive protocol, revealing that this setting can overestimate genuine transferability.
        \end{tablenotes} 
    \end{threeparttable}
\end{table}

\begin{table*}[!t]
    \centering
    \begin{threeparttable}
        \caption{ASR (\%) of transferable attacks under stricter protocols designed to reduce distortion-driven and objective-driven overestimation. Higher is better.}
        \label{tab_performance_transferable_attacks_nonUT16}
        \setlength{\tabcolsep}{0.8mm}
        \begin{tabular}{l|ccccc|ccccc|ccccc}
            \toprule
            \multirow{2}{*}{Attack}
            &\multicolumn{5}{c|}{Untargeted, $\epsilon=4/255$}
            &\multicolumn{5}{c|}{Targeted, $\epsilon=16/255$}
            &\multicolumn{5}{c}{Targeted, $\epsilon=4/255$}\\
            &ResNet-50&ConvNeXt-B&ViT-B&Swin-B&Avg.
            &ResNet-50&ConvNeXt-B&ViT-B&Swin-B&Avg.
            &ResNet-50&ConvNeXt-B&ViT-B&Swin-B&Avg.\\
            \midrule
            PGD&\cellcolor{gray!20}100.0&3.4&1.4&3.6&2.8&\cellcolor{gray!20}99.4&0.0&0.0&0.0&0.0&\cellcolor{gray!20}99.4&0.0&0.0&0.0&0.0\\
            FAP&\cellcolor{gray!20}99.6&9.2&2.6&5.2&5.7&\cellcolor{gray!20}73.0&0.2&0.0&0.0&0.1&\cellcolor{gray!20}94.0&0.0&0.0&0.0&0.0\\
            BFA&\cellcolor{gray!20}99.8&13.2&2.0&5.6&6.9&\cellcolor{gray!20}27.0&0.8&0.6&0.4&0.6&\cellcolor{gray!20}23.4&0.0&0.0&0.0&0.0\\
            P2FA&\cellcolor{gray!20}94.2&10.8&1.2&4.4&5.5&\cellcolor{gray!20}64.4&1.4&0.0&0.6&0.7&\cellcolor{gray!20}37.8&0.0&0.0&0.0&0.0\\
            AWT&\cellcolor{gray!20}99.4&14.4&3.0&7.0&8.1&\cellcolor{gray!20}13.6&0.0&0.0&0.0&0.0&\cellcolor{gray!20}0.0&0.0&0.0&0.0&0.0\\
            OPS&\cellcolor{gray!20}96.4&21.8&8.2&13.0&14.3&\cellcolor{gray!20}100.0&20.2&10.6&4.4&11.7&\cellcolor{gray!20}95.0&0.8&0.0&0.0&0.3\\
            MEF&\cellcolor{gray!20}100.0&15.6&2.8&7.4&8.6&\cellcolor{gray!20}61.8&3.2&0.4&1.2&1.6&\cellcolor{gray!20}27.2&0.0&0.0&0.0&0.0\\
            \bottomrule
        \end{tabular}
        \begin{tablenotes}
            \item Gray cells indicate WASR. Avg. denotes the average TASR, excluding WASR.
            \item The small-budget untargeted protocol is our primary protocol. Targeted transfer with $\epsilon=16/255$ controls for untargeted-objective bias, while targeted transfer with $\epsilon=4/255$ serves as an overly strict stress test. Across these protocols, TASR drops substantially even when WASR remains high.
        \end{tablenotes} 
    \end{threeparttable}
\end{table*}

\noindent \textbf{Attack configurations.}
We use setting-specific attack parameters, summarized in \Cref{tab_attack_settings}, because the three settings differ in defense strength and optimization cost. For classifier transfer, all attacks follow the TransferAttack setting. For purification transfer, all adaptive attacks in \Cref{tab_RW_adaptive_attacks}, including BPDA+EOT, are evaluated under a unified protocol when permitted by the original algorithm: they share the same transfer-only setting, clean-correct subset, downstream classifier, DDIM surrogate checkpoint and configuration, PGD backbone, perturbation budget, and iteration budget. We keep only intrinsic method-specific differences: DiffPGD does not use EOT, DiffHammer replaces EOT with five multi-pass purification evaluations, and DiffBreak uses an LPIPS constraint instead of an $\ell_\infty$ constraint. For DiffBreak, we set the LPIPS bound to $0.0258$, matching PGDTransfer's LPIPS distance under the same setting. For VQA transfer, PGDTransfer follows the VQA attack setting in \Cref{tab_attack_settings} across all visual encoders. Additional adaptive-attack configurations are summarized in \Cref{tab_full_adaptive_settings} of Appendix~\ref{app_full_settings}.

\begin{table}[!t]
    \centering
    \begin{threeparttable}
        \caption{ASR (\%) of transferable attacks between non-robust and robust models with the same architecture under untargeted transfer with $\ell_\infty$ budget $\epsilon=4/255$. Higher is better.}
        \label{tab_performance_transferable_attacks_architecture}
        \setlength{\tabcolsep}{2.5mm}
        \begin{tabular}{lccc}
            \toprule
            \multirow{2}{*}{Attack}&Surrogate&\multicolumn{2}{c}{Target Model}\\
            &Model&ResNet-50&ResNet-50 (Robust)\\
            \midrule
            \multirow{2}{*}{PGD}&ResNet-50&\cellcolor{gray!20}100.0&0.4\\
            &ResNet-50 (Robust)&5.6&\cellcolor{gray!20}32.2\\
            \cmidrule{1-4}
            FAP&ResNet-50&\cellcolor{gray!20}99.6&0.6\\
            &ResNet-50 (Robust)&5.4&\cellcolor{gray!20}30.0\\
            \cmidrule{1-4}
            BFA&ResNet-50&\cellcolor{gray!20}99.8&1.6\\
            &ResNet-50 (Robust)&2.8&\cellcolor{gray!20}8.6\\
            \cmidrule{1-4}
            P2FA&ResNet-50&\cellcolor{gray!20}94.2&0.6\\
            &ResNet-50 (Robust)&1.4&\cellcolor{gray!20}2.0\\
            \cmidrule{1-4}
            AWT&ResNet-50&\cellcolor{gray!20}99.4&2.0\\
            &ResNet-50 (Robust)&5.0&\cellcolor{gray!20}15.8\\
            \cmidrule{1-4}
            OPS&ResNet-50&\cellcolor{gray!20}96.4&2.2\\
            &ResNet-50 (Robust)&2.2&\cellcolor{gray!20}13.4\\
            \cmidrule{1-4}
            MEF&ResNet-50&\cellcolor{gray!20}100.0&1.2\\
            &ResNet-50 (Robust)&6.4&\cellcolor{gray!20}18.6\\
            \bottomrule
        \end{tabular}
        \begin{tablenotes}
            \item Gray cells indicate WASR.
            \item Transfer remains weak in both directions despite identical architecture, showing that architecture alone does not explain transferability.
        \end{tablenotes} 
    \end{threeparttable}
\end{table}

\subsection{Validating the Evaluation Protocols}
\label{sec_protocol_validation}

This section validates the protocol concerns discussed in \Cref{sec_protocol_problem} and the stricter evaluation design introduced in \Cref{sec_stricter_protocols}. The goal is not to identify the strongest transferable attack on non-robust classifiers, but to show that commonly used protocols can overestimate genuine transferability and that stricter protocols better separate transferable adversarial directions from distortion-induced degradation.

\Cref{tab_performance_transferable_attacks_ut16} first reports results under the commonly used protocol: untargeted transfer with $\epsilon=16/255$. The first seven rows follow the standard practice of using a non-robust ResNet-50 surrogate. Under this setting, PGD shows limited transferability, with only 6.2\% average TASR, while designed transferable attacks achieve substantially higher TASR, up to 71.9\%. However, the same protocol can produce a very different conclusion when the surrogate is changed: PGD crafted on a robust Swin-B surrogate reaches 78.5\% average TASR against non-robust target classifiers. This result supports the first concern in \Cref{sec_protocol_problem}: high untargeted TASR under a large budget can depend strongly on the protocol and surrogate choice, and therefore may overestimate genuine transferable adversarial directions.

\Cref{tab_performance_transferable_attacks_nonUT16} evaluates the same attacks under stricter protocols. Under the primary protocol, i.e., untargeted transfer with $\epsilon=4/255$, all methods show much lower TASR, with the strongest average TASR reduced to 14.3\%. Under targeted transfer, TASR further decreases: even with $\epsilon=16/255$, most methods remain near zero, and the strongest average TASR is 11.7\%. With both targeted transfer and $\epsilon=4/255$, transferability is almost completely suppressed. Importantly, WASR remains high in many cases, indicating that the attacks still succeed on their surrogates. The drop therefore mainly reflects reduced transferability rather than failed optimization.

Finally, \Cref{tab_performance_transferable_attacks_architecture} tests whether architecture similarity alone explains transferability. When the surrogate and target share the same ResNet-50 architecture but differ in robustness optimization, all attacks transfer poorly in both directions. This result supports the second concern in \Cref{sec_protocol_problem}: architectural similarity is not sufficient for transferability when the models differ in robust behavior.

Together, these results support the two pitfalls identified in \Cref{sec_protocol_problem}. Common protocols can overestimate genuine transferability, while stricter protocols from \Cref{sec_stricter_protocols} better separate distortion-driven degradation, objective-driven effects, architecture-driven transfer, and robustness-induced shared vulnerability.

\begin{figure*}[!t]
    \centering
    \includegraphics[width=\linewidth]{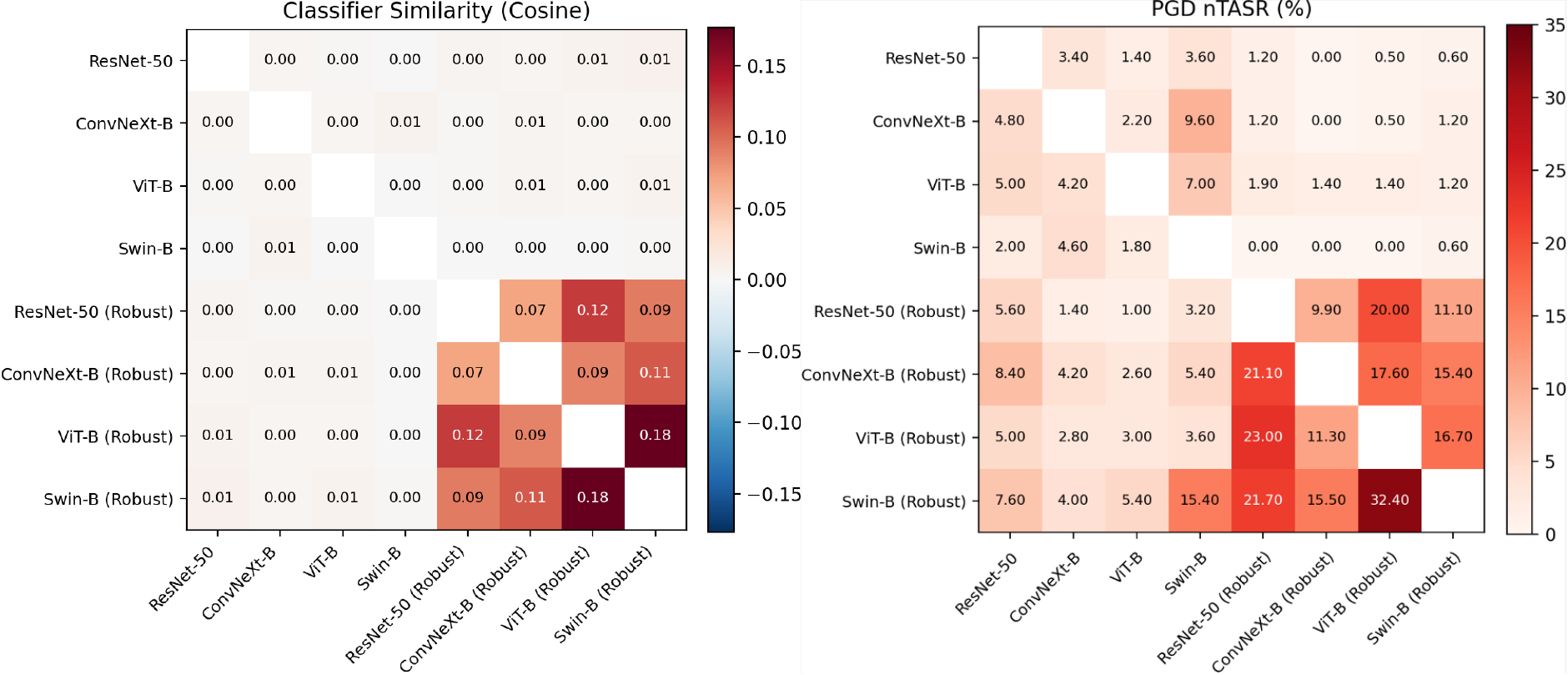}
    \caption{AdvSM similarity and nTASR among non-robust and robust classifiers. The left matrix reports AdvSM cosine similarity, and the right matrix reports PGD nTASR under untargeted transfer with $\ell_\infty$ budget $\epsilon=4/255$. Robust classifiers form a positively aligned AdvSM block, which corresponds to higher nTASR within the robust-classifier block.}
    \label{fig_matrix_classifiers}
\end{figure*}

\begin{table}[!t]
    \centering
    \begin{threeparttable}
        \caption{ASR (\%) of PGD and OPS across robust classifiers under untargeted transfer with $\ell_\infty$ budget $\epsilon=4/255$. Higher is better.}
        \label{tab_performance_transferable_attacks_robust}
        \setlength{\tabcolsep}{1mm}
        \begin{tabular}{lc|cccc|c}
            \toprule
            \multirow{2}{*}{Attack}&Surrogate&\multicolumn{5}{c}{Target Model}\\
            &Model&ResNet-50&ConvNeXt-B&ViT-B&Swin-B&Avg.\\
            \midrule
            \multirow{4}{*}{PGD}&ResNet-50&\cellcolor{gray!20}32.2&1.4&8.4&3.6&\multirow{4}{*}{5.8}\\
            &ConvNeXt-B&6.8&\cellcolor{gray!20}14.2&7.4&5.0&\\
            &ViT-B&7.4&1.6&\cellcolor{gray!20}42.0&5.4&\\
            &Swin-B&7.0&2.2&13.6&\cellcolor{gray!20}32.4&\\
            \midrule
            \multirow{4}{*}{OPS}&ResNet-50&\cellcolor{gray!20}13.4&0.8&9.0&3.2&\multirow{4}{*}{5.9}\\
            &ConvNeXt-B&8.8&\cellcolor{gray!20}4.4&9.6&4.6&\\
            &ViT-B&6.8&0.8&\cellcolor{gray!20}17.4&3.6&\\
            &Swin-B&8.2&1.6&13.8&\cellcolor{gray!20}8.4&\\
            \bottomrule
        \end{tabular}
        \begin{tablenotes}
            \item Gray cells indicate WASR. Avg. denotes the average TASR, excluding WASR. All surrogate and target classifiers are adversarially trained.
            \item PGD achieves a TASR comparable to OPS, the strongest transferable baseline, suggesting that transferability among robust classifiers is driven more by shared sensitivity than by specialized transferable-attack design.
        \end{tablenotes}
    \end{threeparttable}
\end{table}


\emph{In the following sections, we report results under our primary evaluation protocol, i.e., untargeted transfer with $\epsilon=4/255$. Results under other protocols, including untargeted transfer with $\epsilon=16/255$ and targeted transfer with $\epsilon=16/255$, are provided in Appendix~\ref{app_other_protocol}.}

\subsection{Transferability across Classifiers}
\label{sec_transfer_classifiers}

This section evaluates whether adversarially trained, i.e., robust, classifiers exhibit shared adversarial sensitivity, and whether such similarity leads to transferable adversarial examples even when the attack itself is simple. In this setting, PGDTransfer is equivalent to standard PGD.

\Cref{fig_matrix_classifiers} shows that AdvSM similarity is near zero for non-robust--non-robust and non-robust--robust pairs, but becomes consistently positive among robust classifiers. The nTASR matrix follows the same structure: transfer remains weak outside the robust-model block, but becomes stronger within it. This result supports the claim in \Cref{sec_advsm_at}: robustness optimization can align the input regions that remain adversarially sensitive, even across different architectures.

We further compare PGD with OPS, the strongest transferable baseline in \Cref{tab_performance_transferable_attacks_ut16,tab_performance_transferable_attacks_nonUT16}. As shown in \Cref{tab_performance_transferable_attacks_robust}, PGD obtains comparable average TASR to OPS. This is important because PGD is not designed as a transferable attack, whereas OPS explicitly aims to improve transferability. The comparable performance indicates that, once robust classifiers share similar adversarial sensitivity, attack transferability can arise from model alignment rather than from sophisticated transferable-attack mechanisms.

Together, these results support our central claim for adversarially trained classifiers: robustness optimization can induce shared adversarial sensitivity, and this shared sensitivity weakens defense isolation.

\begin{figure}[!t]
    \centering
    \includegraphics[width=\linewidth]{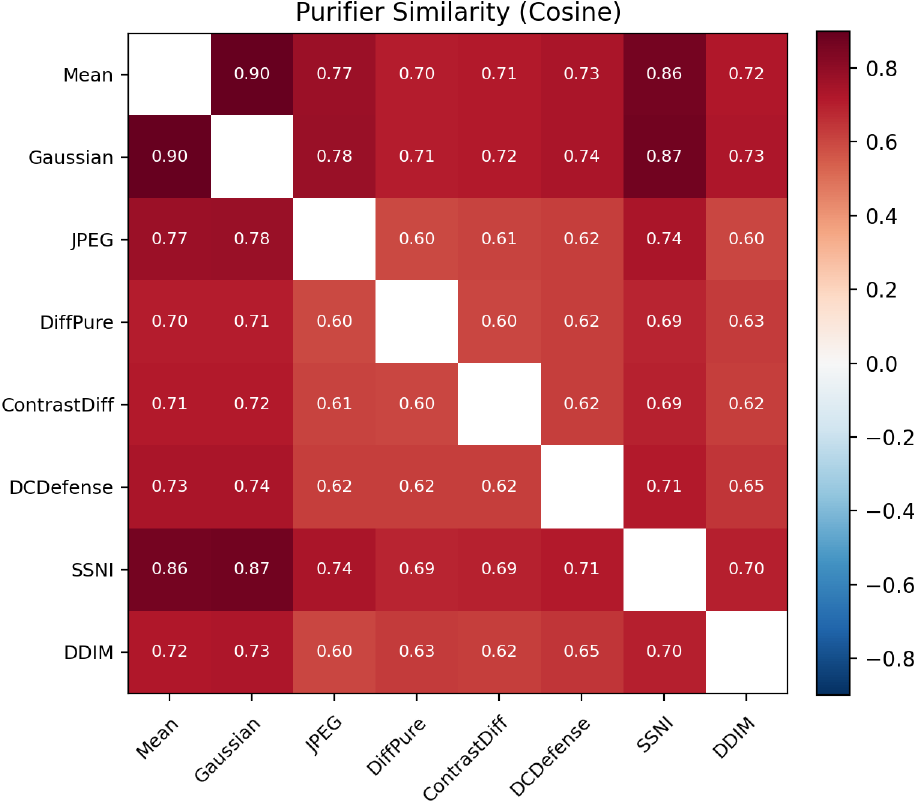}
    \caption{AdvSM similarity among purification-based defenses. Despite different mechanisms, all purifier pairs exhibit positive similarity, indicating shared sensitive regions for adversarial optimization.}
    \label{fig_matrix_purifiers}
\end{figure}

\begin{table*}[!t]
    \centering
    \begin{threeparttable}
        \caption{TASR (\%) of adaptive attacks against purification-based defenses using DDIM as the surrogate purifier under untargeted transfer with $\ell_\infty$ budget $\epsilon=4/255$. Higher is better.}
        \label{tab_performance_adaptive_attacks_ut}
        \setlength{\tabcolsep}{2.2mm}
        \begin{tabular}{lcccccccccc}
            \toprule
            Attack & No Purifier & Mean & Gaussian & JPEG & DiffPure & MimicDiffusion & ContrastDiff & DCDefense & SSNI & Avg. \\
            \midrule
            Clean&100.0&87.0&89.0&89.4&95.2&92.4&95.4&95.6&97.8&93.5\\
            PGD&100.0&31.4&32.8&19.4&6.0&24.8&5.0&6.0&6.2&25.7\\
            \midrule
            BPDA+EOT & \textbf{88.8} & 77.6 & 81.8 & 54.2 & 15.8 & 48.6 & 15.8 & 22.6 & 64.0 & 52.1 \\
            DiffPGD & 69.0 & 86.6 & 88.0 & 67.4 & 49.0 & 64.2 & 44.2 & 50.6 & 75.0 & 66.0 \\
            DiffAttack & 81.6 & 92.4 & 94.2 & 78.4 & 60.0 & 77.0 & 54.6 & 65.6 & 87.0 & 76.8 \\
            DiffHammer & 66.4 & 84.6 & 87.4 & 65.8 & 47.2 & 62.0 & 42.2 & 51.6 & 73.2 & 64.5 \\
            DiffBreak & 49.6 & 68.8 & 69.4 & 63.6 & 53.0 & 50.8 & 47.6 & 56.4 & 63.4 & 58.1 \\
            \midrule
            \textbf{PGDTransfer (ours)} & 87.6 & \textbf{94.2} & \textbf{96.2} & \textbf{82.2} & \textbf{63.6} & \textbf{80.8} & \textbf{58.0} & \textbf{69.6} & \textbf{91.2} & \textbf{80.4} \\
            \bottomrule
        \end{tabular}
        \begin{tablenotes}
            \item Avg. is computed over all target columns, including the undefended classifier baseline.
        \end{tablenotes}
    \end{threeparttable}
\end{table*}

\begin{figure*}[!t]
    \centering
    \includegraphics[width=\linewidth]{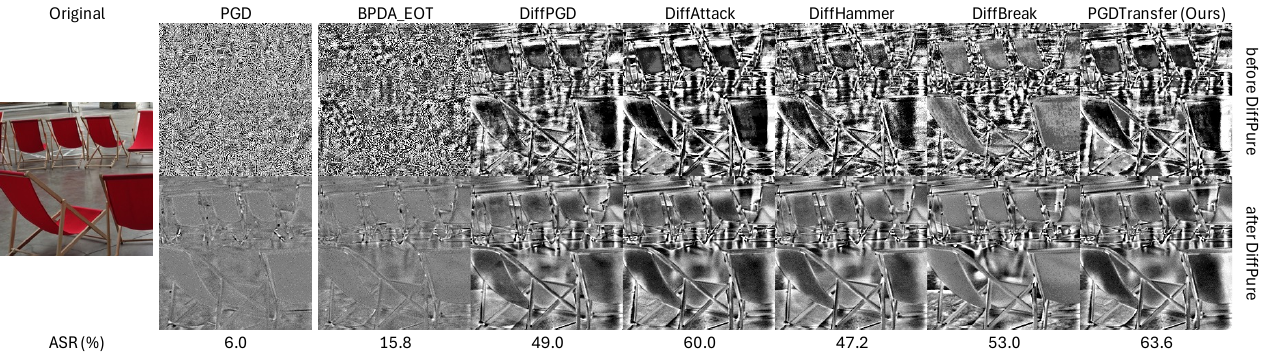}
    \caption{Qualitative comparison of adversarial examples before and after DiffPure purification. The ASR values are measured against DiffPure under the untargeted protocol with $\ell_\infty$ budget $\epsilon=4/255$. Adaptive perturbations can remain effective after purification, showing that purification does not necessarily remove learned adversarial effects.}
    \label{fig_ae}
\end{figure*}

\subsection{Transferability across Adversarial Purifiers}
\label{sec_transfer_purifiers}

This section evaluates transferability across purification-based defenses. These defenses differ substantially in implementation, but share the same high-level objective of removing or suppressing adversarial perturbations before classification. We examine whether they are individually effective against ordinary attacks, whether they exhibit shared AdvSM similarity, and whether one surrogate attack can transfer across the purifier family.

\Cref{tab_performance_adaptive_attacks_ut} first confirms that the evaluated purifiers provide meaningful protection against ordinary non-adaptive PGD. With a fixed non-robust ResNet-50 classifier, the purifiers maintain high clean accuracy, averaging 93.5\%, while reducing untargeted PGD ASR to 25.7\% on average. Therefore, the transfer results below are not caused by trivially weak purification defenses.

\Cref{fig_matrix_purifiers} then shows that the evaluated purifiers nevertheless share strong AdvSM similarity. All purifier pairs have positive similarity, and many pairs are strongly aligned. This supports the observation in \Cref{fig_asm}(d): although these purifiers span filtering, compression, and diffusion-based methods, they still preserve overlapping sensitive regions that remain usable for adversarial optimization.

\begin{table}[!t]
    \centering
    \begin{threeparttable}
        \caption{Time cost per sample of adaptive attacks using DDIM as the surrogate purifier. Lower is better.}
        \label{tab_performance_adaptive_attacks_time}
        \setlength{\tabcolsep}{12.7mm}
        \begin{tabular}{lc}
            \toprule
            Attack & Time Cost \\
            \midrule
            BPDA+EOT&2.2 min\\
            DiffPGD&2.2 min\\
            DiffAttack&2.4 min\\
            DiffHammer$^1$&5.0 min\\
            DiffBreak&2.5 min\\
            \midrule
            PGDTransfer (ours)&2.3 min\\
            \bottomrule
        \end{tabular}
        \begin{tablenotes}
            \item $^1$ DiffHammer does not use EOT, but instead uses multiple evaluations to address stochasticity. These evaluations cannot be efficiently parallelized.
            \item PGDTransfer has a time cost comparable to other adaptive attacks while achieving stronger transfer performance.
        \end{tablenotes} 
    \end{threeparttable}
\end{table}

\Cref{tab_performance_adaptive_attacks_ut} further shows that this shared sensitivity translates into strong attack transferability. Under the primary untargeted protocol with $\epsilon=4/255$, PGDTransfer achieves the highest TASR on every purifier, ranging from 58.0\% to 96.2\%, with an average TASR of 80.4\%. Notably, DDIM surrogate is not only effective against diffusion purifiers, but also transfers strongly to filtering, compression, and undefended targets. These results directly support our claim that, once a suitable surrogate purifier enables effective adversarial optimization, the resulting adversarial examples can transfer across other purifiers within the same robustness family. The comparison with prior adaptive attacks further supports the role of shared purifier sensitivity. PGDTransfer is a simple PGD-style attack with EOT and a DDIM surrogate, yet it outperforms attacks specifically tailored to diffusion-based purification, including DiffPGD, DiffAttack, DiffHammer, and DiffBreak. This suggests that transferability can arise from family-level shared vulnerability rather than from sophisticated attack-specific mechanisms.

\Cref{fig_ae} further illustrates why purification alone is insufficient against adaptive attacks. Using DiffPure as a representative purifier, we compare adversarial images before and after purification. Although DiffPure visibly modifies the inputs, the adversarial effect remains after purification, especially for PGDTransfer. This suggests that adaptive perturbations can be optimized to survive the purification process rather than being simply denoised away.

Finally, \Cref{tab_performance_adaptive_attacks_time} shows that PGDTransfer is computationally practical. It takes 2.3 minutes per sample, comparable to most adaptive baselines. By optimizing through a lightweight DDIM surrogate with only three denoising steps, PGDTransfer avoids direct backpropagation through target purifiers that may be non-differentiable or memory-intensive, while still generating adversarial examples that transfer without target-specific optimization.

Together, these results support our central claim for purification-based defenses. The evaluated purifiers are individually effective against ordinary PGD, yet still share adversarial sensitivity. When this shared sensitivity is exposed by a suitable surrogate, a simple PGD-style adaptive attack can generate adversarial examples that transfer across the evaluated purifier family, weakening defense isolation despite the apparent robustness of individual purifiers.

\begin{figure}[!t]
    \centering
    \includegraphics[width=\linewidth]{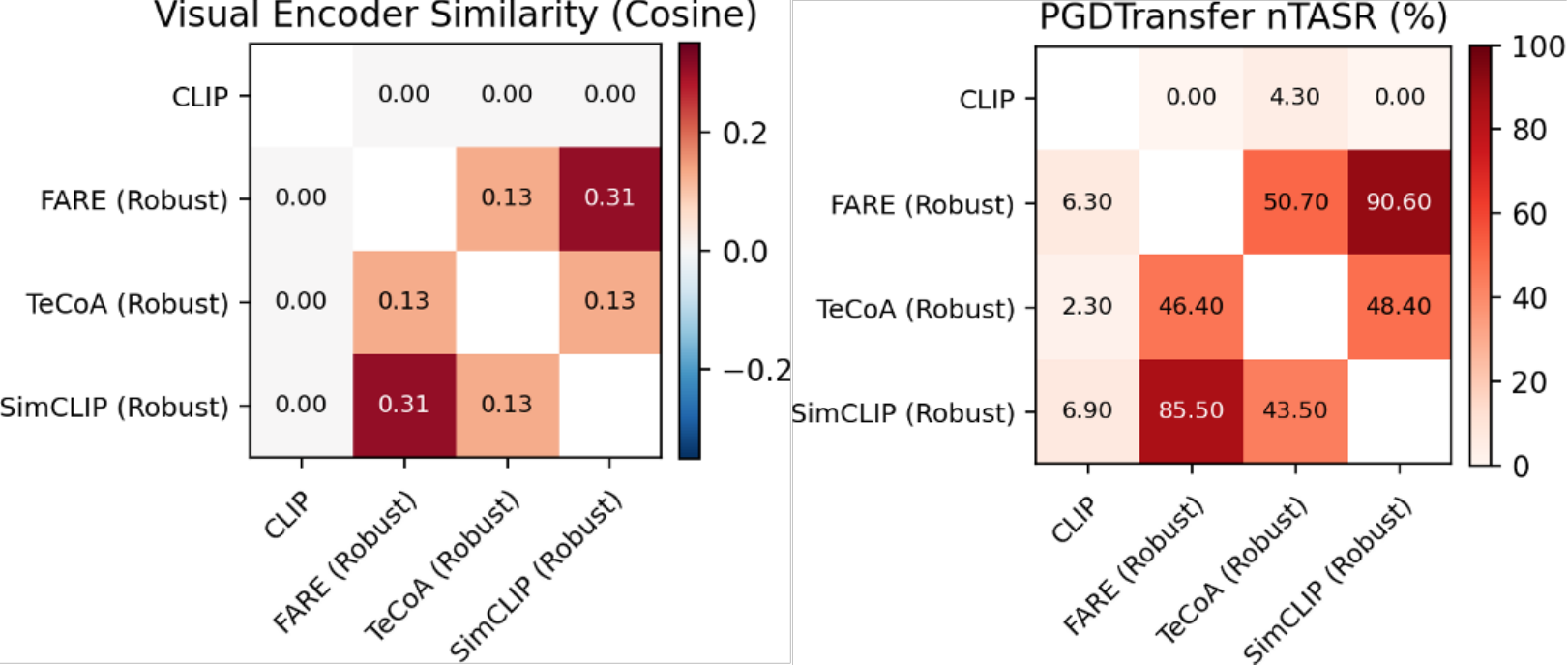}
    \caption{AdvSM similarity and nTASR among non-robust and robust visual encoders. The left matrix reports AdvSM cosine similarity, and the right matrix reports nTASR under untargeted transfer with $\ell_\infty$ budget $\epsilon=4/255$. Robust visual encoders show positive AdvSM similarity, corresponding to stronger transfer within the robust-encoder block.}
    \label{fig_matrix_clips}
\end{figure}

\subsection{Transferability across LVLMs}
\label{sec_transfer_vqa}

This section evaluates whether robustness-induced transferability extends beyond image classification. We use LVLM-based VQA systems that share the same LLaVA-1.5-7B language model and multimodal projector, while varying only the visual encoder to isolate encoder-level transferability.

\Cref{fig_matrix_clips} shows that the non-robust CLIP encoder has near-zero AdvSM similarity with robust encoders, whereas robust encoders exhibit positive similarity with each other. The nTASR matrix follows the same pattern: transfer is weak from CLIP to robust encoders, but stronger within the robust-encoder block. This suggests that adversarial training can align the input regions that remain adversarially sensitive, even when the robust component is embedded inside an LVLM pipeline. 

\Cref{tab_vqa} confirms this pattern. PGD transfer from CLIP to robust-encoder VQA systems is nearly zero, while transfer among robust-encoder systems is consistently higher. Thus, the transferability observed in robust classifiers also appears in LVLMs with robust visual encoders.

Together, these results show that visual encoders optimized toward similar robust behavior can share adversarial sensitivity, weakening defense isolation in VQA systems.

\begin{table}[!t]
    \centering
    \begin{threeparttable}
        \caption{ASR (\%) of PGDTransfer on VQA systems with different visual encoders under untargeted transfer with $\ell_\infty$ budget $\epsilon=4/255$. Higher is better.}
        \label{tab_vqa}
        \setlength{\tabcolsep}{3.2mm}
        \begin{tabular}{lcccc}
            \toprule
            Encoder & CLIP & FARE & TeCoA & SimCLIP \\
            \midrule
            CLIP & \cellcolor{gray!20}35.0 & 0.0 & 0.6 & 0.0 \\
            FARE (Robust) & 2.2 & \cellcolor{gray!20}13.8 & 7.0 & 11.6 \\
            TeCoA (Robust) & 0.8 & 6.4 & \cellcolor{gray!20}13.8 & 6.2 \\
            SimCLIP (Robust) & 2.4 & 11.8 & 6.0 & \cellcolor{gray!20}12.8 \\
            \bottomrule
        \end{tabular}
        \begin{tablenotes}
            \item Gray cells indicate WASR. All systems use the same LLaVA-1.5-7B language model and multimodal projector; only the visual encoder varies.
            \item PGD shows weak transfer from the non-robust CLIP encoder to robust encoders, but stronger transfer among robust encoders, indicating that the classifier-level observation extends to VQA systems.
        \end{tablenotes} 
    \end{threeparttable}
\end{table}

\begin{table}[!t]
    \centering
    \begin{threeparttable}
        \caption{TASR (\%) of PGDTransfer using different surrogate purifiers under untargeted transfer with $\ell_\infty$ budget $\epsilon=4/255$. Higher is better.}
        \label{tab_ablation_surrogate}
        \setlength{\tabcolsep}{4.3mm}
        \begin{tabular}{lccc}
            \toprule
            Target & DDIM (Ours) & Mean & Gaussian\\
            \midrule
            No Purifier&\textbf{87.6}&51.4&72.4\\
            Mean&94.2&\cellcolor{gray!20}\textbf{100.0}&99.8\\
            Gaussian&96.2&\textbf{100.0}&\cellcolor{gray!20}\textbf{100.0}\\
            JPEG&\textbf{82.2}&35.8&39.4\\
            DiffPure&\textbf{63.6}&7.4&7.8\\
            MimicDiffusion&\textbf{80.8}&44.4&54.8\\
            ContrastDiff&\textbf{58.0}&8.8&7.8\\
            DCDefense&\textbf{69.6}&9.8&10.6\\
            SSNI&\textbf{91.2}&13.2&16.2\\
            \midrule
            Avg.&\textbf{80.4}&41.2&45.4\\
            \bottomrule
        \end{tabular}
        \begin{tablenotes}
            \item Gray cells indicate WASR. Avg. is computed over all target columns, including the undefended classifier baseline.
            \item JPEG is non-differentiable, while diffusion-based target purifiers are often memory-intensive for direct gradient backpropagation. We therefore compare feasible differentiable surrogate purifiers.
        \end{tablenotes}
    \end{threeparttable}
\end{table}

\begin{figure*}[!t]
    \centering
    \includegraphics[width=\linewidth]{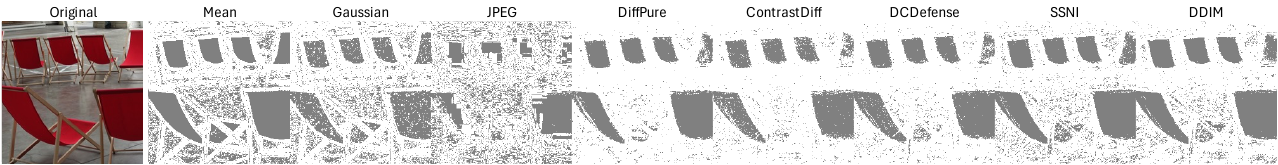}
    \caption{Purifier response maps on a clean input. Filtering- and compression-based purifiers remain closer to identity, whereas diffusion-based purifiers introduce stronger structured changes, explaining why BPDA-style identity-gradient approximations are less suitable for diffusion purification.}
    \label{fig_asm_clean}
\end{figure*}

\subsection{Choosing a Proper Surrogate}
\label{sec_surrogate_selection}

This section validates the surrogate-selection principle in \Cref{sec_proper_surrogate}. Transfer is stronger when surrogate-optimized perturbations remain within sensitive regions preserved by target systems in the same robustness family. If the surrogate exposes too many sensitive regions, perturbations may rely on locations later ignored, smoothed, or purified by the target, making transfer resemble an additional defense pass.

The classifier and LVLM results support this principle. In \Cref{fig_matrix_classifiers,fig_matrix_clips}, transfer is stronger from robust to non-robust systems than in reverse. This asymmetry is consistent with \Cref{fig_asm}(b,c): robust classifiers and visual encoders expose compact semantic sensitivity regions, whereas non-robust ones produce broader, noisier maps. Perturbations optimized on compact robust regions can affect non-robust targets, but those optimized on broader non-robust regions are more likely to be ignored by robust targets.

The purification results show the same pattern. \Cref{tab_ablation_surrogate} compares DDIM, Mean, and Gaussian as surrogate purifiers. Mean and Gaussian transfer well to filtering-based targets but poorly to stronger diffusion-based purifiers. In contrast, DDIM achieves the best average TASR of 80.4\% and is the strongest surrogate for most targets. This matches \Cref{fig_asm}(d): weaker purifiers leave many sensitive regions available, whereas stronger purifiers suppress or smooth more regions. Thus, perturbations optimized through a weak purifier may rely on regions later removed by stronger targets, while a compact surrogate reduces this secondary-cancellation effect.


\section{Additional Insights from AdvSMs}
\label{app_other_advsm}

Beyond explaining transferability, AdvSMs also reveal additional properties of purification-based defenses. These observations help interpret how purifiers suppress perturbations and why different adaptive attacks behave differently.

\subsection{Purification Rarely Reverses Perturbations}
\label{app_reverse_perturbations}

\Cref{fig_asm}(d) suggests that adversarial purification rarely removes perturbations by reversing them back to the original clean input. Instead, many locations become smooth regions with stable one-directional responses (black in the maps). In our AdvSM definition, these regions indicate that local perturbations are suppressed or redirected. In contrast, fewer regions become fully purified (gray in the maps), which suggests that purification often weakens adversarial perturbations by smoothing them into consistent directions rather than by restoring the original response.

\subsection{Why BPDA Fails on Diffusion Purifiers}
\label{app_bpda_diffusion}

\Cref{fig_asm_clean} visualizes purifier response maps on clean inputs. Filtering- and compression-based purifiers, such as Mean, Gaussian, and JPEG, largely preserve input structure and remain close to identity, making BPDA's identity approximation relatively effective. In contrast, diffusion-based purifiers introduce stronger structured changes even on clean inputs. Their forward process is therefore far from identity, so BPDA's identity-gradient approximation no longer matches the actual purifier behavior. This explains why BPDA remains effective against simpler filtering- or compression-based purifiers, but becomes much less effective against diffusion-based purification.

\section{Conclusion}
\label{sec_conclusion}

This work shows that adversarial robustness optimization can introduce a shared-vulnerability dimension across defenses. When systems are optimized toward similar robust behavior, their remaining adversarial sensitivity can become aligned, allowing adversarial examples crafted on one suitable surrogate to transfer across other systems in the same robustness family. This weakens defense isolation: even if each defense appears robust in isolation, breaching one representative system may expose the broader family without target-side gradients, queries, or adaptation.

The broader insight is that robustness is not only an individual property of a model or pipeline, but also a relational property among defenses developed under similar objectives. Future defenses should therefore optimize not only for standalone robustness, but also for vulnerability diversity. A more secure direction is to reduce shared sensitivity across defenses, for example through gradient or AdvSM decoupling, diversified robust feature reliance, and transfer-only adaptive evaluation.
\section*{Acknowledgments}

This work was partially supported by JSPS KAKENHI Grant JP24H00732, by JST CREST Grants JPMJCR20D3 and JPMJCR2562 including AIP challenge program, and by JST K Program Grant JPMJKP24C2 Japan.
\section*{Ethical Considerations}

This work studies adversarial transferability among robustness-optimized defenses. The main ethical risk is dual use: the proposed analysis and PGDTransfer could help improve adaptive attacks against deployed ML systems. We mitigate this risk by evaluating only public research models, public datasets, and offline experimental settings. We do not attack live services, collect private user data, or interact with systems outside our control. The evaluated datasets are standard public benchmarks, and our experiments do not involve human subjects or personally identifiable information.

The purpose of this work is to improve security evaluation. Our results show that defenses should be evaluated not only by standalone robustness, but also by defense isolation under transfer-only attacks. We will submit an anonymized artifact for review to support reproducibility (\emph{\url{https://github.com/azrealwang/AdvSM}}), and will publicly release the code and configurations upon publication.
{
    \bibliographystyle{IEEEtran}
    \bibliography{references}
}

%

\appendices

\section{Full Experimental Settings}
\label{app_full_settings}

This appendix summarizes the experimental settings used throughout this work. 
\Cref{tab_full_eval_settings} lists the shared datasets, system configurations, and AdvSM settings. 
\Cref{tab_full_adaptive_settings} details the unified adaptive-attack configurations used for purification transfer. For adaptive attacks, all methods are evaluated under the same transfer-only threat model, downstream ResNet-50 classifier, DDIM surrogate checkpoint and denoising configuration, clean-correct subset, and optimization budget whenever permitted by the original algorithm. Method-specific differences are preserved only when required by the original attack.

\begin{table}[!t]
    \centering
    \begin{threeparttable}
        \caption{Shared evaluation settings and pointers to detailed configurations.}
        \label{tab_full_eval_settings}
        \setlength{\tabcolsep}{2.0mm}
        \renewcommand{\arraystretch}{1.12}
        \begin{tabular}{p{2.3cm}p{5.4cm}}
            \toprule
            \textbf{Item} & \textbf{Setting} \\
            \midrule
            Threat model & Transfer-only; no target gradients, queries, or target-specific adaptation. \\
            Clean subset & 500 clean-correct samples for attack evaluation unless otherwise specified. \\
            AdvSM subset & 100 clean-correct samples. \\
            Classification data & NIPS 2017 adversarial defense challenge dataset. \\
            Purification data & NIPS 2017 adversarial defense challenge dataset. \\
            VQA data & VQAv2. \\
            Classifier transfer & TransferAttack implementation; single-surrogate setting. \\
            Purifier pipeline & Fixed non-robust ResNet-50 downstream classifier; only the purifier varies. \\
            VQA pipeline & Fixed LLaVA-1.5-7B language model and multimodal projector; only the visual encoder varies. \\
            Classifier / VQA AdvSM & Gradient threshold $\tau=10^{-5}$. \\
            Purifier AdvSM & Random sign perturbation $\epsilon=16/255$; response threshold $\theta=2/255$; perturbation samples $M=5$; purification runs $N=10$. \\
            Purifier settings & Listed in \Cref{tab_settings_purification}. \\
            Main attack settings & Listed in \Cref{tab_attack_settings}. \\
            \bottomrule
        \end{tabular}
    \end{threeparttable}
\end{table}

\begin{table*}[!t]
    \centering
    \begin{threeparttable}
        \caption{Unified adaptive-attack configurations for purification transfer.}
        \label{tab_full_adaptive_settings}
        \setlength{\tabcolsep}{1mm}
        \renewcommand{\arraystretch}{1.12}
        \begin{tabular}{lccccp{8.8cm}}
            \toprule
            \textbf{Attack} 
            & \textbf{Objective} 
            & \textbf{Constraint} 
            & \textbf{Iter.} 
            & \textbf{Stochastic handling} 
            & \textbf{Method-specific component} \\
            \midrule
            BPDA+EOT
            & Untargeted
            & $\ell_\infty=4/255$
            & 40
            & EOT $=5$
            & Uses the identity-gradient approximation in the backward pass. \\
            
            DiffPGD
            & Untargeted
            & $\ell_\infty=4/255$
            & 40
            & None
            & Does not use EOT by original algorithm design. \\
            
            DiffAttack
            & Untargeted
            & $\ell_\infty=4/255$
            & 40
            & EOT $=5$
            & Preserves its original deviated-reconstruction loss and modified DDIM logic. \\
            
            DiffHammer
            & Untargeted
            & $\ell_\infty=4/255$
            & 40
            & Multi-pass $=5$
            & Replaces EOT with five multi-pass purification evaluations following the original design. \\
            
            DiffBreak
            & Untargeted
            & LPIPS $=0.0258$
            & 40
            & EOT $=5$
            & Uses an LPIPS constraint instead of an $\ell_\infty$ constraint; the bound matches the LPIPS distance of PGDTransfer under the same setting. \\
            
            PGDTransfer
            & Untargeted
            & $\ell_\infty=4/255$
            & 40
            & EOT $=5$
            & Uses the standard task loss through the DDIM surrogate. \\
            \bottomrule
        \end{tabular}
        \begin{tablenotes}
            \item All attacks use PGD as the optimization backbone with step size $\alpha=1/255$, the same DDIM surrogate purifier, the same DDIM checkpoint and denoising configuration, and the same downstream ResNet-50 classifier. The DDIM surrogate uses time steps $=150$ and denoising steps $=3$.
            \item The targeted adaptive-transfer setting uses the same unified configuration with targeted objective and $\ell_\infty=16/255$, except that each method keeps its original stochastic-handling mechanism.
        \end{tablenotes}
    \end{threeparttable}
\end{table*}

\begin{table*}[!t]
    \centering
    \begin{threeparttable}
        \caption{TASR (\%) of adaptive attacks against purification-based defenses using DDIM as the surrogate purifier under targeted transfer with $\ell_\infty$ budget $\epsilon=16/255$. Higher is better.}
        \label{tab_performance_adaptive_attacks_t}
        \setlength{\tabcolsep}{2.2mm}
        \begin{tabular}{lcccccccccc}
            \toprule
            Attack & No Purifier & Mean & Gaussian & JPEG & DiffPure & MimicDiffusion & ContrastDiff & DCDefense & SSNI & Avg. \\
            \midrule
            Clean&100.0&87.0&89.0&89.4&95.2&92.4&95.4&95.6&97.8&93.5\\
            PGD&100.0&3.2&4.6&2.2&0.0&1.6&5.8&0.8&1.2&13.3\\
            \midrule
            BPDA+EOT&54.8&45.0&53.2&34.8&8.3&8.0&7.2&15.6&49.4&30.7\\
            DiffPGD&64.6&69.6&73.0&53.8&34.0&39.2&32.0&42.4&72.8&53.5\\
            \midrule
            \textbf{PGDTransfer (ours)}& \textbf{93.0} & \textbf{93.2} & \textbf{96.0} & \textbf{84.2} & \textbf{66.0} & \textbf{73.8} & \textbf{62.0} & \textbf{75.6} & \textbf{96.6} & \textbf{82.3} \\
            \bottomrule
        \end{tabular}
        \begin{tablenotes}
            \item Avg. is computed over all target columns, including the undefended classifier baseline.
        \end{tablenotes}
    \end{threeparttable}
\end{table*}

\begin{table}[!t]
    \centering
    \begin{threeparttable}
        \caption{TASR (\%) of PGDTransfer under different $\ell_\infty$ budgets. Higher is better.}
        \label{tab_ablation_eps}
        \setlength{\tabcolsep}{5.8mm}
        \begin{tabular}{lccc}
            \toprule
            $\epsilon$&4/255&8/255&16/255\\
            \midrule
            No Purifier&87.6&99.6&100.0\\
            Mean&94.2&99.8&100.0\\
            Gaussian&96.2&100.0&100.0\\
            JPEG&82.2&99.6&100.0\\
            DiffPure&63.6&97.8&100.0\\
            MimicDiffusion&80.8&97.4&100.0\\
            ContrastDiff&58.0&97.0&100.0\\
            DCDefense&69.6&98.2&100.0\\
            SSNI&91.2&99.6&100.0\\
            Avg.&80.4&98.8&100.0\\
            \bottomrule
        \end{tabular}
        \begin{tablenotes}
            \item Even at $\epsilon=4/255$, PGDTransfer transfers strongly across purifiers. Under the larger budget $\epsilon=16/255$, all targets reach 100\% TASR.
        \end{tablenotes}
    \end{threeparttable}
\end{table}

\begin{table}[!t]
    \centering
    \begin{threeparttable}
        \caption{ASR (\%) of PGD and OPS across robust classifiers under untargeted transfer with $\ell_\infty$ budget $\epsilon=16/255$. Higher is better.}
        \label{tab_performance_transferable_attacks_robust_16}
        \setlength{\tabcolsep}{1mm}
        \begin{tabular}{lc|cccc|c}
            \toprule
            \multirow{2}{*}
            {Attack}&Surrogate&\multicolumn{5}{c}{Target Model}\\
            &Model&ResNet-50&ConvNeXt-B&ViT-B&Swin-B&Avg.\\
            \midrule
            \multirow{4}{*}{PGD}&ResNet-50&\cellcolor{gray!20}88.2&19.8&51.4&39.6&\multirow{4}{*}{37.2}\\
            &ConvNeXt-B&36.6&\cellcolor{gray!20}63.0&36.8&37.8&\\
            &ViT-B&41.4&16.4&\cellcolor{gray!20}95.2&56.8&\\
            &Swin-B&34.8&19.8&54.6&\cellcolor{gray!20}92.4&\\
            \midrule
            \multirow{4}{*}{OPS}&ResNet-50&\cellcolor{gray!20}68.0&19.0&60.6&40.4&\multirow{4}{*}{48.5}\\
            &ConvNeXt-B&58.6&\cellcolor{gray!20}48.2&65.0&51.4&\\
            &ViT-B&52.8&21.0&\cellcolor{gray!20}82.0&51.0&\\
            &Swin-B&57.4&29.0&75.2&\cellcolor{gray!20}70.0&\\
            \bottomrule
        \end{tabular}
        \begin{tablenotes}
            \item Gray cells indicate WASR. Avg. denotes the average TASR, excluding WASR. All surrogate and target classifiers are adversarially trained.
        \end{tablenotes}
    \end{threeparttable}
\end{table}

\begin{table}[!t]
    \centering
    \begin{threeparttable}
        \caption{ASR (\%) of PGDTransfer on VQA systems with different visual encoders under untargeted transfer with $\ell_\infty$ budget $\epsilon=16/255$. Higher is better.}
        \label{tab_vqa_16}
        \setlength{\tabcolsep}{3.2mm}
        \begin{tabular}{lcccc}
            \toprule
            Encoder & CLIP & FARE & TeCoA & SimCLIP \\
            \midrule
            CLIP & \cellcolor{gray!20}39.6 & 0.4 & 0.8 & 0.4 \\
            FARE (Robust) & 31.0 & \cellcolor{gray!20}34.6 & 25.0 & 33.4 \\
            TeCoA (Robust) & 28.2 & 34.6 & \cellcolor{gray!20}35.8 & 34.8 \\
            SimCLIP (Robust) & 29.8 & 32.6 & 23.8 & \cellcolor{gray!20}33.2 \\
            \bottomrule
        \end{tabular}
        \begin{tablenotes}
            \item Gray cells indicate WASR. All systems use the same LLaVA-1.5-7B language model and multimodal projector; only the visual encoder varies.
        \end{tablenotes} 
    \end{threeparttable}
\end{table}

\section{Additional Results under Other Protocols}
\label{app_other_protocol}

The main paper uses untargeted transfer with $\epsilon=4/255$ as the primary protocol because it better reduces distortion-driven overestimation. 
This appendix reports complementary results under larger-budget and targeted settings: targeted purification transfer at $\epsilon=16/255$ (\Cref{tab_performance_adaptive_attacks_t}), purification budget ablations (\Cref{tab_ablation_eps}), robust-classifier transfer at $\epsilon=16/255$ (\Cref{tab_performance_transferable_attacks_robust_16}), and VQA transfer at $\epsilon=16/255$ (\Cref{tab_vqa_16}). 
These results are not used as primary evidence for genuine transferability, but they characterize how transfer changes when the perturbation budget is enlarged or the objective is made more restrictive.

Overall, the complementary results support the same interpretation under different protocols. 
First, targeted purification transfer with $\epsilon=16/255$ is more restrictive than untargeted transfer, yet PGDTransfer still achieves strong transfer when DDIM is used as the surrogate (\Cref{tab_performance_adaptive_attacks_t}). 
Second, the purification budget ablation shows that the larger untargeted budget is highly permissive: PGDTransfer reaches 100\% TASR across all evaluated purifier settings at $\epsilon=16/255$ (\Cref{tab_ablation_eps}). 
Third, robust-classifier transfer at $\epsilon=16/255$ remains stronger within the robustness-optimized family, consistent with the main-protocol results (\Cref{tab_performance_transferable_attacks_robust_16}). 
Finally, VQA transfer at $\epsilon=16/255$ shows the same qualitative pattern: transfer is weak from non-robust CLIP to robust encoders but stronger among robust visual encoders (\Cref{tab_vqa_16}).

\section{Ablation Study on PGDTransfer}
\label{app_ablation_pgdtransfer}

We further ablate two implementation choices in PGDTransfer: the number of EOT samples and the number of denoising steps in the DDIM surrogate. These results are not used as main evidence, but they help explain why our default setting balances attack effectiveness and computational cost.

\Cref{tab_ablation_eot} shows the effect of EOT. Using a single stochastic forward pass, which corresponds to the DiffPGD-style setting, already gives non-trivial transferability. Increasing EOT to 5 substantially improves TASR across all evaluated targets, especially on the undefended classifier, JPEG, and DiffPure. Further increasing EOT to 10 gives only marginal additional gains but greatly increases runtime because the extra stochastic evaluations cannot be efficiently parallelized under our memory constraints. We therefore use EOT=5 as the default setting.

\Cref{tab_ablation_deno_steps} shows the effect of the DDIM surrogate strength. With only one denoising step, the surrogate is too weak and leaves many perturbation directions available, leading to lower transferability. Increasing the number of denoising steps improves transfer, but the gain saturates after three steps. Using five steps slightly improves some targets but increases runtime substantially. We therefore use three denoising steps as the default surrogate setting, which exposes a compact set of adversarially useful regions while keeping optimization practical.

\begin{table}[!t]
    \centering
    \begin{threeparttable}
        \caption{TASR (\%) of PGDTransfer under different EOT settings. Higher ASR indicates stronger attacks.}
        \label{tab_ablation_eot}
        \setlength{\tabcolsep}{2.1mm}
        \begin{tabular}{lccccc}
            \toprule
            EOT&Time&No Purifier&Mean&JPEG&DiffPure\\
            \midrule
            1 (DiffPGD)&2.2 min&69.0&86.6&67.4&49.0\\
            5 (ours)&2.3 min&87.6&94.2&82.2&63.6\\
            10$^1$&21.4 min&89.2&95.4&84.8&67.4\\
            \bottomrule
        \end{tabular}
        \begin{tablenotes}
            \item $^1$ When EOT $>5$, it cannot be efficiently parallelized due to memory constraints.
        \end{tablenotes} 
    \end{threeparttable}
\end{table}

\begin{table}[!t]
    \centering
    \begin{threeparttable}
        \caption{TASR (\%) of PGDTransfer under different denoising steps in the DDIM surrogate. Higher ASR indicates stronger attacks.}
        \label{tab_ablation_deno_steps}
        \setlength{\tabcolsep}{1.7mm}
        \begin{tabular}{lccccc}
            \toprule
            Denoising Steps&Time&No Purifier&Mean&JPEG&DiffPure\\
            \midrule
            1&1.9 min&67.4&93.2&77.4&57.6\\
            2&2.0 min&81.8&94.0&80.2&63.0\\
            3 (ours)&2.3 min&87.6&94.2&82.2&63.6\\
            5$^1$&10.4 min&91.0&94.8&82.6&64.0\\
            \bottomrule
        \end{tabular}
        \begin{tablenotes}
            \item $^1$ When denoising steps $>3$, the attack cannot be efficiently parallelized due to memory constraints.
        \end{tablenotes} 
    \end{threeparttable}
\end{table}

\section{Related Works}
\label{sec_related_work}

\noindent \textbf{Robustness-optimized defenses.}
Adversarial training (\Cref{tab_RW_at}) improves robustness by optimizing models on adversarial examples, and has been studied for both image classifiers and visual encoders in LVLMs~\cite{salman2020adversarially,mo2022adversarial,liu2025comprehensive,schlarmann2024robust,mao2023understanding,hossain2026sim}. Adversarial purification (\Cref{tab_RW_purification}) instead preprocesses inputs before prediction, using filtering, compression, GANs, VAEs, EBMs, or diffusion models~\cite{lee1980digital,jain1989fundamentals,wallace1991jpeg,samangouei2018defense,jin2019ape,yoon2021adversarial,hill2021stochastic,shi2021online,yu2024purify,nie2022diffusion,lee2023robust,song2024mimicdiffusion,bai2024diffusion,chen2024diffilter,li2025adbm,pei2025divide,sun2025sample,wang2025iterative,zollicoffer2025lorid,park2025adversarial}. Unlike detection-based defenses, these two types of defenses aim to preserve correct prediction under perturbation, and therefore fall within the robustness-optimization scope studied in this work.

\noindent \textbf{Transferable attacks.}
Transferable attacks (\Cref{tab_RW_transferable_attacks}) aim to craft adversarial examples on a surrogate model that remain effective on unseen targets. Recent methods improve transferability through Bayesian substitutes, frequency-aware perturbations, feature-space objectives, pixel-to-feature optimization, adversarial weight tuning, hypothesis-space augmentation, and flatness-based optimization~\cite{li2023making,wang2024boosting,wang2024improving,liu2025pixel2feature,chen2025enhancing,guo2025boosting,qiu2026boosting}. These attacks mainly evaluate transfer among non-robust classifiers. In contrast, we study whether robustness optimization itself induces transferability among defenses within the same robustness family.

\noindent \textbf{Adaptive attacks on purification defenses.}
Adaptive attacks (\Cref{tab_RW_adaptive_attacks}) evaluate defended pipelines rather than undefended classifiers. BPDA+EOT approximates non-differentiable preprocessing with differentiable backward passes~\cite{athalye2018obfuscated}, while recent attacks against diffusion purifiers use DDIM surrogates, EOT, modified diffusion states, EM-style objectives, or perceptual constraints~\cite{xue2023diffusion,kang2023diffattack,wang2024diffhammer,kassis2025diffbreak}. Our PGDTransfer follows this adaptive setting but keeps the attack mechanism simple. Its purpose is to test whether shared adversarial sensitivity among robustness-optimized defenses is sufficient to induce transferability, rather than to rely on increasingly specialized attack design.

\end{document}